\documentclass[aps,pra,twocolumn,showpacs,superscriptaddress]{revtex4} 
\usepackage{graphicx}
\usepackage{epstopdf}
\usepackage{amsmath}
\usepackage{amssymb}
\usepackage{epstopdf}
\usepackage{amsfonts}
\usepackage{dcolumn}
\usepackage{bigints}
\usepackage{xcolor}
\usepackage{orcidlink}
\usepackage[normalem]{ulem}

\usepackage{dirtytalk}
\usepackage{csquotes}

\MakeOuterQuote{"}

\newcommand{\flr}[1]{\left\lfloor #1\right\rfloor}

\usepackage{color, soul}

\newcommand{\lp }{\left(}
\newcommand{\rp }{\right)}

\newcommand{\lb }{\left[}
\newcommand{\rb }{\right]}

\newcommand{\parder}[1]{\frac{\partial}{\partial #1}}

\newcommand{\intinf}{\int_{-\infty}^\infty}

\begin{document}

\title{Efficient Determination of eigenenergies and eigenstates of $N$ ($N=3$--$4$) identical 1D bosons and fermions under external harmonic confinement}
\author{J. D. Norris}
\email{Jacob.D.Norris-1@ou.edu}
\address{Homer L. Dodge Department of Physics and Astronomy,
  The University of Oklahoma,
  440 W. Brooks Street,
  Norman,
Oklahoma 73019, USA}
\address{Center for Quantum Research and Technology,
  The University of Oklahoma,
  440 W. Brooks Street,
  Norman,
Oklahoma 73019, USA}
\author{D. Blume\,\orcidlink{0000-0001-7381-5698}}
\email{doerte.blume-1@ou.edu}
    \address{Homer L. Dodge Department of Physics and Astronomy,
  The University of Oklahoma,
  440 W. Brooks Street,
  Norman,
Oklahoma 73019, USA}
\address{Center for Quantum Research and Technology,
  The University of Oklahoma,
  440 W. Brooks Street,
  Norman,
Oklahoma 73019, USA}
    
\begin{abstract}
Few-atom systems play an important role in understanding the transition from few- to many-body quantum behaviors. This work introduces a new approach for determining the energy spectra and eigenstates of small harmonically trapped single-component Bose and Fermi gases with additive two-body zero-range interactions in one spatial dimension. The interactions for bosons are the usual $\delta$-function interactions while those for fermions are $\delta$-function interactions that contain derivative operators. Details of the derivation and benchmarks of the numerical scheme are presented. Extensions to other systems are discussed.  
\end{abstract}
\maketitle

\section{Introduction}

Effectively low-dimensional quantum gases have received a great deal of attention over the past few decades due to their enhanced quantum fluctuations and unique quantum correlations~\cite{RevModPhys.85.1633, MISTAKIDIS20231, 10.1088/1361-6633/ab3a80, 10.1051/jp4:2004116001}. 
Experimentally, effectively low-dimensional quantum gases can be realized by tightly confining the motion of the particles along one or two spatial dimensions, resulting in effectively two-dimensional and effectively one-dimensional systems, respectively~\cite{PhysRevLett.91.250402, PhysRevLett.87.160406, PhysRevLett.99.040402, doi:10.1126/science.1100700, PhysRevLett.104.153203}. If the chemical potential is sufficiently low, i.e., smaller than the excitations in the tight confinement direction(s), the system properties can frequently be described faithfully by assuming that the motion along the tight confining directions is frozen out~\cite{PhysRevLett.85.3745, PhysRevLett.86.5413}. Intriguingly, the tight confinement does provide one with a “knob” for changing the interaction strength in the reduced dimensionality space by means of so-called confinement induced resonances~\cite{PhysRevLett.81.938, PhysRevLett.91.163201, PhysRevLett.104.153203, PhysRevA.64.012706, PhysRevLett.92.133202, PhysRevLett.100.170404}. 

This work is devoted to small strictly one-dimensional bosonic systems and strictly one-dimensional fermionic systems under external harmonic confinement. Starting with the Lippmann-Schwinger equation~\cite{PhysRevA.76.033611}, we derive eigenvalue equations for three and four identical bosons as well as three identical fermions under external harmonic confinement. We introduce a novel computationally efficient and  numerically stable scheme for solving the eigenvalue equations, yielding the energy spectra as a function of the interaction strength over a fairly broad energy range (20 or more converged eigenenergies). For the four-boson system, these are the first calculations of this type. An important result of our work is that the Lippmann-Schwinger equation approach for identical fermions yields stable results. This indicates that the odd-parity interaction potential, which contains a regularization operator~\cite{PhysRevA.70.042709, PhysRevLett.82.2536, CHEON1998111, PhysRevLett.95.230404, HaraldGrosse_2004}, yields mathematically well-defined expressions for systems with more than two particles.

Our study complements earlier work on strictly one-dimensional few-atom systems, including studies of two-component Fermi gases and multi-component Bose and Fermi gases under external harmonic confinement~\cite{busch,PhysRevA.91.023620, doi:10.1126/sciadv.1500197, slxg-7dnm, PhysRevA.86.052122, PhysRevA.91.013620, PhysRevA.70.042709, PhysRevA.91.043634, PhysRevA.78.013629}. Motivated by the Bose-Fermi mapping theorem~\cite{10.1063/1.1703687, GIRARDEAU20043, Yukalov_2005, PhysRevLett.82.2536}, small single-component Bose and Fermi gases have received a great deal of attention. The mapping, which applies to any interaction strength, has been investigated most extensively in the  Tonks-Girardeau regime, where strongly-interacting single-component bosons can be mapped to weakly-interacting single-component fermions. Analogously, strongly-interacting single-component fermions can be mapped to weakly-interacting single-component bosons. While interacting three-dimensional single-component fermions tend to suffer from comparatively large losses~\cite{PhysRevLett.90.053201, PhysRevLett.95.230401, https://doi.org/10.1038/nphys3670}, they may be stabilized by effectively low-dimensional confinement~\cite{PhysRevA.96.030701, PhysRevLett.125.263402, marcum2020suppressionthreebodylossnear}. Our work adds to the comparatively small body of literature that is treating interacting one-dimensional single-component  Fermi gases directly~\cite{10.1088/1367-2630/abb386, PhysRevLett.130.253401}, without resorting to the Bose-Fermi duality. Future work will extend our framework to two-dimensional single-component fermions, which have been predicted to support super-Efimov states~\cite{PhysRevLett.110.235301, PhysRevA.92.020504, Gridnev_2014, Volosniev_2014},  and to three-dimensional single-component fermions~\cite{PhysRevLett.94.023202, PhysRevA.70.042709, PhysRevA.77.043611, PhysRevLett.96.013201}, which feature modified threshold laws~\cite{PhysRevA.65.010705, PhysRevLett.90.053202, PhysRevA.105.013308}.

    The remainder of this paper is structured as follows. Section~\ref{sec_systems} introduces the system Hamiltonian under study and the theoretical framework employed to solve the corresponding time-independent Schr\"odinger equation. Section~\ref{sec_numerics} details our numerical scheme to evaluate the integrals that enter into the eigenvalue equation. Section~\ref{sec_spectra} presents and interprets the resulting energy spectra. Finally, Sec.~\ref{sec_conclusion} concludes. The  derivation of the eigenvalue equation for three identical fermions is relegated to Appendix~\ref{appendixfff}.
    
\section{Systems under Study and Theoretical Framework}
\label{sec_systems}
\subsection{System Hamiltonian and Lippmann-Schwinger equation}
We consider $N$ identical one-dimensional particles of mass $m$ with position coordinates $\mathbf{x}$, where \(\mathbf{x}=(x_1,\dots,x_N)\), under  external harmonic confinement with angular trapping frequency \(\omega\). The total system Hamiltonian $\hat{H}_{\text{tot}}$ is given by
        \begin{eqnarray}
            \hat{H}_{\text{tot}}=\hat{H}_{\text{tot,NI}}+\hat{V}_{\text{int}}, \label{genham}
        \end{eqnarray}
    where $\hat{H}_{\text{tot,NI}}$ is the sum of $N$ single-particle harmonic oscillator Hamiltonians $\hat{H}_{\text{ho}}(x_j,m)$,
    \begin{eqnarray}
        \hat{H}_{\text{tot,NI}}=\sum_{j=1}^N \hat{H}_{\text{ho}}(x_j,m)
    \end{eqnarray} 
    with
    \begin{eqnarray}
        \hat{H}_{\text{ho}}(x_j,m) =
        -\frac{\hbar^2}{2m}\frac{\partial^2}{\partial x_j^2}+\frac{1}{2}m\omega^2 x_j^2.
    \end{eqnarray} 
    The form of the interaction potential
    \(\hat{V}_{\text{int}}\) depends on the system under consideration. For $N$ identical bosons and $N$ identical fermions, we use a sum of two-body zero-range potentials $V_+(x_{jk})$, where $x_{jk}=x_j-x_k$, with even-parity coupling constant $g_+$ and a sum of two-body zero-range potentials $V_-(x_{jk})$ with odd-parity coupling constant $g_-$, respectively,
    \begin{eqnarray}\label{intpot}
        \hat{V}_{\text{int}}=
        \sum_{j=1}^{N-1}\sum_{k>j}^N
        V_{\pm}(x_{jk}),
    \end{eqnarray}
    where 
    \begin{eqnarray}
        V_{+}(x_{jk})=g_{+}\delta(x_{jk}) \label{vp}
    \end{eqnarray}
    and 
    \begin{eqnarray}
        V_{-}(x_{jk})=-g_{-}\left[\frac{\partial}{\partial x_{jk}}\delta(x_{jk}) \right]\frac{\partial}{\partial x_{jk}}.\label{vm}
    \end{eqnarray}
  The derivative operators are needed  since functions that are odd under the exchange of  particles $j$ and $k$ vanish at $x_{jk}=0$; the derivative operator extracts the, in general, finite value of the partial derivative of the wavefunction. For $\omega=0$, the Hamiltonian for $V_+(x_{jk})$ coincides with the famous Lieb-Liniger Hamiltonian~\cite{PhysRev.130.1605}.
    
    The form of $V_-(x_{jk})$, Eq.~(\ref{vm}), can be obtained from the expression~\cite{PhysRevA.70.042709}
    \begin{eqnarray}\label{vm_alter}
       V_-(x_{jk})=g_{-}\overleftarrow{\frac{\partial}{\partial x_{jk}}}\delta\lp x_{jk}\rp \overrightarrow{\frac{\partial}{\partial x_{jk}}}
    \end{eqnarray}
    by observing that the matrix elements for the potentials given in Eqs.~(\ref{vm}) and (\ref{vm_alter}) for 
    functions \(f(x_{jk})\) and \( g(x_{jk})\), which should be differentiable, square integrable,  and  obey \(f(0)=0\) or \(\partial^2 g(x_{jk})/\partial x_{jk}^2|_{x_{jk}=0}=0\) [if the former, \(\partial^2 g(x_{jk})/\partial x_{jk}^2|_{x_{jk}=0}\) should be zero or finite; if the latter, \(f(0)\) should be zero or finite], are the same. 
    Specifically, we find 
        \begin{eqnarray}
            \int_{-\infty}^\infty f(x) \overleftarrow{\frac{\partial}{\partial x}}\delta\lp x\rp\overrightarrow{\frac{\partial}{\partial x}} g(x)dx  =
            \nonumber \\ 
            \int_{-\infty}^\infty \frac{\partial f(x)}{\partial x}\delta(x) \frac{\partial g(x)}{\partial x} dx, 
        \end{eqnarray}
        which evaluates to $[\partial f(x)/\partial x]_{x=0} [\partial g(x)/\partial x]_{x=0}$,
    and
        \begin{align}
        \label{vm_aux1}
            \int_{-\infty}^\infty f(x)\lb -\frac{\partial \delta(x)}{\partial x}\frac{\partial g(x) }{\partial x}\rb dx&= \nonumber \\ -f(x) \frac{\partial g(z)}{\partial x} \delta(x)\bigg|_{-\infty}^{\infty}
            + \lb\frac{\partial}{\partial x}\lp f(x)\frac{\partial g(x)}{\partial x}\rp\rb_{x=0} . 
        \end{align}
        Using that the surface term [first term on the right-hand side of Eq.~(\ref{vm_aux1})] vanishes since $f(x)$ is square integrable (i.e., it vanishes at $x=\pm \infty$) and using the chain rule to expand the second term, we find that Eq.~(\ref{vm_aux1}) reduces to $[\partial f(x)/\partial x]_{x=0} [\partial g(x)/\partial x]_{x=0}$, where we used that $f(0)[\partial^2g(x)/\partial x^2]_{x=0}=0$ by our assumption.
        The condition  \(f(0)=0\) and \(\partial^2 g(x_{jk})/\partial x_{jk}^2|_{x_{jk}=0}\) finite or zero, or \(\partial^2 g(x_{jk})/\partial x_{jk}^2|_{x_{jk}=0}=0\) and \(f(0)\) finite or zero is fulfilled for all  matrix elements encountered in this work that involve the odd-parity pseudopotential. For example,  
    $f(0)=0$ is fulfilled for the relative two-body odd-parity Green's function evaluated at zero. 
    Furthermore, \(\partial^2 g(x_{jk})/\partial x_{jk}^2|_{x_{jk}=0}=0\) is fulfilled for the relative two-fermion wavefunction because the fermionic wavefunction is antisymmetric under the exchange of the position coordinates, implying that its  first and second derivatives are even and odd, respectively.

    To proceed, 
    we change from single-particle coordinates $\mathbf{x}$ to $N-1$ relative  coordinates 
    $\mathbf{z}$, where $\mathbf{z}=(z_{1},\dots,z_{N-1})$, and the center-of-mass coordinate $Z$,
     $Z=N^{-1}\sum_{j=1}^N x_j$.
     The masses associated with the relative coordinates are denoted by $\mu_1,\dots,\mu_{N-1}$ while the mass associated with the center-of-mass coordinate is denoted by $M$, $M=Nm$. Since the center-of-mass degree of freedom is not impacted by the interactions, the eigenstates and eigenenergies of the center-of-mass Hamiltonian coincide with those of the harmonic oscillator Hamiltonian $\hat{H}_{\text{ho}}(Z,M)$. 
     
     In what follows, we focus on solving the time-independent Schr\"odinger equation
     \begin{eqnarray}\label{tise}
         \hat{H} \psi(\mathbf{z})=E \psi(\mathbf{z})
     \end{eqnarray}
     for the relative Hamiltonian $\hat{H}$,
     \begin{eqnarray}\label{relham}
\hat{H}=\hat{H}_{\text{NI}}+\hat{V}_{\text{int}},
\end{eqnarray}
where
\begin{eqnarray}
    \hat{H}_{\text{NI}}=\sum_{j=1}^{N-1} \hat{H}_{\text{ho}}(z_j,\mu_j).
\end{eqnarray}
Introducing the Green's function $G(E,\mathbf{z},\mathbf{z}')$,
the eigenstates $\psi(\mathbf{z})$ can be written as~\cite{PhysRevA.76.033611}
\begin{eqnarray}
\label{eq_ls_wavefunction}
        \psi(\mathbf{z})=-\int   G(E,\mathbf{z},\mathbf{z}') \hat{V}_{\text{int}} \psi(\mathbf{z}')d\mathbf{z}',\label{lse}
        \end{eqnarray}
where
\begin{eqnarray}
\label{greenfuncnbody}
G(E,\mathbf{z},\mathbf{z}')=
            \sum_{\mathbf{n}}\frac{\Phi_{\mathbf{n}}^*(\mathbf{z}')\Phi_{\mathbf{n}}(\mathbf{z})}{E_{\mathbf{n}}-E}.
        \end{eqnarray}
        Here, $\mathbf{n}$ is a collective quantum number,
        $\mathbf{n}=(n_1,\dots,n_{N-1})$, with each of the $n_j$ quantum numbers ($j=1,\dots,N-1$)  running independently from 0 to infinity. 
        The $\Phi_{\mathbf{n}}(\mathbf{z})$ are eigenstates of $\hat{H}_{\text{NI}}$ with eigenenergy $E_{\mathbf{n}}$,
\begin{eqnarray}
    \Phi_{\mathbf{n}}(\mathbf{z})=\prod_{j=1}^{N-1} \varphi_{n_j}(z_j,\mu_j)
\end{eqnarray}
and
\begin{eqnarray}
E_{\mathbf{n}}=\sum_{j=1}^{N-1} E_{n_j}.
\end{eqnarray}
The $\varphi_{n_j}(z_j,\mu_j)$, 
\begin{eqnarray}
    \varphi_{n_j}(z_j,\mu_j)= \nonumber \\
    \lp\frac{\mu_j \omega}{\pi \hbar}\rp^{\frac{1}{4}}\frac{1}{\sqrt{2^{n_j} n_j !}} H_{n_j}\lp \sqrt{\frac{\mu_j \omega}{\hbar}} z_j\rp \exp \left(-\frac{\mu_j \omega}{2\hbar} z_j^2\right),
\end{eqnarray}
in turn, are eigenstates of $\hat{H}_{\text{ho}}(z_j,\mu_j)$ with eigenenergy 
$E_{n_j}= (n_j+1/2)\hbar \omega$, where $n_j=0,1,\dots$. 

        \subsection{Eigenvalue equations}
A key advantage of applying the Green's function or Lippmann-Schwinger equation  approach to systems with zero-range interactions is that one of the integrals on the right-hand side of Eq.~(\ref{lse}) can be performed straightforwardly. For the three- and four-body systems, we are thus left with, respectively, one and two non-trivial integrations.  This work considers systems that consist of three identical bosons, three identical fermions, and four identical bosons.
     We use the Jacobi coordinates  
     $\mathbf{z}=(z_{12},z_{12,3})$ for the three-body systems and  $\mathbf{z}=(z_{12},z_{34},z_{12,34})$ for the four-body system,
     where  
    \begin{eqnarray}\label{jacobi}
    z_{ij}&=&x_i-x_j, \\ 
    z_{ij,k}&=&\frac{2}{\sqrt{3}}\lp \frac{x_i+x_j}{2}-x_k\rp, \mbox{ and }\\ z_{ij,k\ell}&=&\frac{1}{\sqrt{2}}\lp x_i+x_j-x_k-x_\ell\rp.
    \end{eqnarray}
    The associated Jacobi masses are denoted by $\mu_{ij}$, $\mu_{ij,k}$, and  $\mu_{ij,k\ell}$, respectively, where  $\mu_{ij}=\mu_{ij,k}=\mu_{ij,k\ell}=m/2$.
    
    Since the relative parity operator $\hat{\Pi}$, which sends $\mathbf{z}$ to $-\mathbf{z}$, commutes with the relative Hamiltonian $\hat{H}$,
    the energy spectra can be divided into the positive relative parity sector ($\Pi=+1$) and the negative relative parity sector ($\Pi=-1$). For the bosonic and fermionic systems, the energetically lowest-lying eigenstate has, respectively, positive and negative relative parity. Correspondingly, we consider $\Pi=+1$ for the BBB and BBBB systems and $\Pi=-1$ for the FFF system.
      Employing similar techniques to those used in Ref.~\cite{PhysRevA.76.033611, PhysRevA.86.042702, PhysRevLett.111.045302, PhysRevLett.102.160401, PhysRevA.82.023619}, we derive eigenvalue equations for the BBB and BBBB systems, which yield the interaction strengths \(g_{+}\) for a given relative energy \(E\). Appendix~\ref{appendixfff} provides a detailed  derivation of the eigenequation for the FFF system, highlighting subtleties that arise due to the derivative operators that enter through $\hat{V}_{\text{int}}$.

For the BBB system  ($\Pi=+1$), the eigenvalue equation reads~\cite{10.1088/0953-4075/47/6/065303} 
\begin{eqnarray}
\label{goaleqbbb}
            \frac{1}{g_{+}}a_\ell = 
            \sum_{k=0, 2,\dots}
            A_{\ell k}^{\text{(3B)}}(E-E_\ell)
            a_{k},
        \end{eqnarray}
        where $\ell$ is equal to $0,2,\dots$ and the  superscript ``3B'' stands for ``BBB.'' Interpreting $1/g_+$ as the eigenvalue of the energy-dependent matrix elements $A_{\ell k}^{\text{(3B)}}(E-E_\ell)$,
        the coefficients $a_\ell$ can be thought of as components of the eigenvector. 
        The matrix elements $A_{\ell k}^{\text{(3B)}}(E-E_\ell)$ are expressed in terms of the limiting behavior of the two-body even-parity Green's function $G^{(\text{2B})}(E-E_\ell,z,z')$ and the integral $I_{\ell k}^{\text{(3B)}}(E-E_\ell)$, 
        \begin{eqnarray}
            \label{eq_eigen_bbb}
            A_{\ell k}^{(\text{3B})}(E-E_\ell)= \nonumber \\
            -G^{(\text{2B})}\lp E-E_\ell,~0,~0\rp\delta_{\ell k} -2
            I_{\ell k}^{\text{(3B)}}(E-E_\ell) ,
        \end{eqnarray}
        where~\cite{busch, PhysRevA.70.042709, PhysRevA.86.042702}
        \begin{align}
        \label{eq_greenstwo}
             G^{(\text{2B})}(E-E_\ell,0,0)=\frac{1}{2\hbar\omega a_{\text{ho}}} \frac{\Gamma \left(-\frac{(E-E_\ell)}{2\hbar\omega}+\frac{1}{4} \right)}{\Gamma \left(-\frac{(E-E_\ell)}{2\hbar\omega}+\frac{3}{4}\right)}
        \end{align}
        and~\cite{10.1088/0953-4075/47/6/065303} 
        \begin{eqnarray}
        \label{eq_integral_capi}
        I_{\ell k}^{\text{(3B)}}(E-E_\ell)=\nonumber \\
        \int_{-\infty}^{\infty}
        G^{(\text{2B})}\lp E-E_\ell,0,~\frac{\sqrt{3}}{2}u\rp
        \varphi_\ell^*\lp -\frac{u}{2},\frac{m}{2}\rp\varphi_k\lp u,\frac{m}{2}\rp  du.\nonumber \\
        \end{eqnarray}
        In Eq.~(\ref{eq_greenstwo}), $a_{\text{ho}}$ denotes the harmonic oscillator length, $a_{\text{ho}}=\sqrt{2\hbar/(m\omega)}$. 
         The form and evaluation of the two-body Green's function $G^{(\text{2B})}(E-E_\ell,z,z')$ for two identical bosons is discussed in detail in Sec.~\ref{sec_numerics}. In practice, the infinite sum in Eq.~(\ref{goaleqbbb}) is truncated at a finite cutoff, thereby reducing the problem to the diagonalization of a finite square matrix. The dominant computational complexity lies  in evaluating the  integral $I_{\ell k}^{\text{(3B)}}(E-E_\ell)$~\cite{PhysRevLett.111.045302, PhysRevA.86.042702}.  Section~\ref{sec_numerics} presents an efficient computational scheme that builds on recursion relationships~\cite{son2024derivationrecursiveformulasintegrals}. The coefficients $a_0,a_2,\dots$, obtained for a fixed $E$, determine the  eigenstates $\psi(z_{12},z_{12,3})$,
        \begin{eqnarray}
            \psi(z_{12},z_{12,3}) = \nonumber \\
            \hat{\mathcal{S}} \lb \sum_{\ell=0,2,\dots} a_\ell  N_{\ell} G^{\text{(2B)}}(E-E_{\ell}, z_{12}, 0) \varphi_\ell (z_{12,3},\mu_{12,3})\rb, \nonumber \\
        \end{eqnarray}
            where $N_{\ell}$ denotes a normalization constant ($N_{\ell}$ is chosen to be real and positive)  and 
 $\hat{\mathcal{S}}$ denotes the symmetrization operator of the three-boson system.
        
        We note that Eq.~(\ref{eq_eigen_bbb}) only yields eigenstates whose energies are impacted by the interaction potential $\hat{V}_{\text{int}}$. In addition, there exist eigenstates that are  not impacted by $\hat{V}_{\text{int}}$ (for three harmonically trapped three-dimensional bosons, this was highlighted in Ref.~\cite{PhysRevLett.97.150401}). Their energies are equal to 
        $(n_1+n_2+1)\hbar \omega$, where $n_1=1,3,\dots$ and $n_2=1,3,\dots$.
        The corresponding eigenstates are given by
        \begin{eqnarray}
            \label{eq_eigenstate_bbb_unshifted}
            \psi(z_{12},z_{12,3})=\hat{\cal{S}} \left[ \varphi_{n_1}(z_{12},\mu_{12})\varphi_{n_2}(z_{12,3},\mu_{12,3})\right].
        \end{eqnarray}
        Since these eigenstates go to zero when two particles approach each other, their energy is independent of $g_+$. The energies of these ``unshifted'' states are not shown in the energy spectra discussed in  Sec.~\ref{sec_spectra}.

    For the FFF system in the relative parity sector 
    $\Pi=-1$, the eigenequation is (see Appendix~\ref{appendixfff}) 
        \begin{align}\label{goaleqfff}
            \frac{1}{g_{-}}a_{\ell} &=\sum_{k=0,2,\dots} A^{\text{(3F)}}_{\ell k}(E-E_{\ell})a_k,
        \end{align}
        where \(\ell\) is equal to \(0,2,\dots\) and where the superscript ``3F'' stands for ``FFF,''
        \begin{widetext}
        \begin{align}
            A^{\text{(3F)}}_{\ell k}(E-E_{\ell}) = 
            \nonumber \\ 
            -\left[\frac{\partial^2 G^{(\text{2F})}(E-E_{\ell},z,z')}{\partial z \partial z'}\right]_{z=z'=0}\delta_{k,\ell} - 
            \frac{4}{\sqrt{3}}\lb \varphi_{\ell}^*\left(0,\frac{m}{2}\right)\varphi_k\left(0,\frac{m}{2}\right)\rb 
            -\frac{2}{\sqrt{3}}J^{\text{(3F)}}_{\ell k}(E-E_{\ell})
            +\frac{4}{\sqrt{3}}K^{\text{(3F)}}_{\ell k}(E-E_{\ell}),
        \end{align}
        \begin{align}
        \label{eq_integral_capj}
            J^{\text{(3F)}}_{\ell k}(E-E_\ell)= 
            \intinf 
            \left[ \frac{\partial G^{(\text{2F})}\lp E-E_\ell,z,\frac{\sqrt{3}}{2}u\rp}{\partial z} \right]_{z=0}
            \varphi_{\ell}^*\lp -\frac{1}{2}u,\frac{m}{2}\rp \frac{\partial \varphi_k(u,\frac{m}{2})}{\partial u}
            du,
        \end{align}
        and
            \begin{align}
            \label{eq_integral_capk}
                K^{\text{(3F)}}_{\ell k}(E-E_\ell)= 
                \intinf \left[\frac{\partial G^{(\text{2F})}\lp E-E_\ell,z,\frac{\sqrt{3}}{2}u \rp}{\partial z} \right]_{z=0}
                \left[
                \frac{\partial \varphi_\ell^{*}\lp z,\frac{m}{2}\rp }{\partial z}\right]_{z=-u/2} \varphi_k\left(u,\frac{m}{2}\right)
                du.
            \end{align}
            \end{widetext}
            The Green's function $G^{(\text{2F})}(E-E_\ell,z,z')$ for two
            identical fermions is defined in Sec.~\ref{sec_numerics}. The limiting behavior can be written as (this follows from Ref.~\cite{PhysRevA.70.042709}) 
            \begin{align}
                \left[\frac{\partial^2 G^{(\text{2F})}(E-E_\ell,z,z')}{\partial z \partial z'}\right]_{z=z'=0} &
                = \nonumber \\
                -\frac{1}{(\hbar\omega a_{\text{ho}}^2)^2G^{(\text{2B})}(E-E_\ell,0,0)}.
            \end{align}
            The inverse relationship between the relative two-boson Green's function and the second derivative of the relative two-fermion Green's function is closely  related to the Bose-Fermi duality~\cite{PhysRevA.70.042709}.

    For the BBBB system ($\Pi=+1$), the eigenvalue equation reads 
            \begin{eqnarray}\label{goaleqbbbb}
                \frac{1}{g_{+}}b_{st}=\sum_{k,\ell=0,2,\dots} B^{\text{(4B)}}_{stk \ell}(E-E_s-E_t)b_{k \ell},
            \end{eqnarray}
        where \(s\) and \(t\) independently take the values \(0,2,\dots\). Similarly to above, interpreting \(1/g_{+}\) as the eigenvalue of the energy-dependent matrix elements \(B^{(\text{4B})}_{stk \ell}(E-E_s-E_t)\), the coefficients \(b_{st}\) can be thought of as components of the eigenvector. The matrix elements \(B^{\text{(4B)}}_{stk \ell}(E-E_s-E_t)\) are expressed in terms of  \(G^{(\text{2B})}(E-E_s-E_t,0,0)\) [see Eq. \eqref{eq_greenstwo}], the term \(J^{\text{(4B)}}_{stk \ell}(E-E_s-E_t)\), and the integral \(I^{\text{(4B)}}_{stk \ell}(E-E_s-E_t)\),
            \begin{eqnarray}
                B^{\text{(4B)}}_{stk \ell}(E-E_s-E_t)= 
                \nonumber \\
                -G^{(\text{2B})}(E-E_s-E_t,0,0)\delta_{s k}\delta_{t \ell}- \nonumber \\J^{\text{(4B)}}_{stk \ell}(E-E_s-E_t)- 
                4 I^{\text{(4B)}}_{stk \ell}(E-E_s-E_t),
            \end{eqnarray}
            where
        \begin{eqnarray}
            J^{\text{(4B)}}_{stk \ell}(E-E_s-E_t)=
            \delta_{t \ell} \frac{\varphi_s^*\left(0,\frac{m}{2}\right)\varphi_k^*\left(0,\frac{m}{2}\right)}{E_k-(E-E_s-E_t)} 
        \end{eqnarray}
        and
            \begin{widetext}
            \begin{eqnarray}
            \label{eq_integral_capi4}
                I^{\text{(4B)}}_{stk \ell}(E-E_s-E_t)= \nonumber \\
                \int_{-\infty}^\infty\int_{-\infty}^\infty G^{(\text{2B})}\lp E-E_s-E_t,0,\frac{u+\sqrt{2}v}{2}\rp \varphi_s^*\lp \frac{u-\sqrt{2}v}{2},\frac{m}{2}\rp \varphi_t^*\lp -\frac{u}{\sqrt{2}},\frac{m}{2}\rp \varphi_k\left(u,\frac{m}{2}\right)\varphi_\ell \left(v,\frac{m}{2}\right) du dv.
            \end{eqnarray}
            \end{widetext}
    It can be seen that the integral $I_{stk \ell}^{\text{(4B)}}(E-E_s-E_t)$ has a similar structure as the integral  $I_{\ell k}^{\text{(3B)}}(E-E_s-E_t)$. 
    The former contains two more harmonic oscillator functions in the integrand than the latter, which is due to the fact that the BBBB system has one more relative coordinate than the BBB system.

\section{Implementation and convergence}
\label{sec_numerics}

This section develops an efficient scheme for constructing the matrix elements $A^{\text{(3B)}}_{\ell k}(E-E_\ell)$, 
$A^{\text{(3F)}}_{\ell k}(E-E_\ell)$, and $B^{\text{(4B)}}_{stk \ell}(E-E_s-E_t)$. 

\subsection{Analytical  considerations}
To evaluate the matrix elements $A^{\text{(3B)}}_{\ell k}(E-E_\ell)$, 
$A^{\text{(3F)}}_{\ell k}(E-E_\ell)$, and $B^{\text{(4B)}}_{stk \ell}(E-E_s-E_t)$, it is useful to introduce a short-hand notation for the single-particle harmonic oscillator functions with mass $m/2$,
\begin{eqnarray}
    \phi_n(u)=\varphi_n\left(u,\frac{m}{2}\right).
\end{eqnarray}
When we write $\phi_n(-u/2)$, we mean that the argument of $\phi_n$ is replaced by $-u/2$ but that the mass factor is unchanged. This implies that $\phi_n(-u/2)=\varphi_n(-u/2,m/2)$ is neither an eigenstate of a harmonic oscillator Hamiltonian nor normalized.

The  recursion relationship~\cite{son2024derivationrecursiveformulasintegrals} 
 \begin{widetext}
        \begin{align}\label{wintegralsrecursion2}
            W_{ijk\ell}=\frac{1}{2}\left( -\sqrt{\frac{i-1}{i}}W_{i-2,jk\ell}+\sqrt{\frac{j}{i}}W_{i-1,j-1,k\ell}+\sqrt{\frac{k}{i}}W_{i-1,j,k-1,\ell}+\sqrt{\frac{\ell}{i}}W_{i-1,jk,\ell-1}\right), 
        \end{align}
        where
        \begin{align}\label{wintegrals2}
            W_{ijk\ell}=\frac{A_{ijk\ell}}{a_{\text{ho}}}\int_{-\infty}^{\infty} H_{i}\left(\frac{u}{a_{\text{ho}}}\right)H_{j}\left(\frac{u}{a_{\text{ho}}}\right)H_{k}\left(\frac{u}{a_{\text{ho}}}\right)H_\ell\left(\frac{u}{a_{\text{ho}}}\right)\exp\left[-2\left(\frac{u}{a_{\text{ho}}}\right)^2\right] du
        \end{align}
         \end{widetext}
    and 
    \begin{eqnarray}    
    A_{ijk\ell}=\frac{1}{\pi \sqrt{2^{i+j+k+\ell}i!j!k!\ell!}},
    \end{eqnarray}
    is a key aspect of our numerical scheme. Importantly, substituting \(x=u/a_{\text{ho}}\) shows that $W_{ijk\ell}$ is independent of the harmonic oscillator length \(a_{\text{ho}}\).
    Both $W_{ijk\ell}$ and $A_{ijk\ell}$ have four subscripts. We  use commas to ``separate'' the subscripts whenever there would exist an ambiguity otherwise (such as, e.g., in $W_{i-2,jk \ell}$).

    We start with the BBB system. Our strategy in what follows is to rewrite the integral $I_{\ell k}^{\text{(3B)}}(E-E_\ell)$, Eq.~(\ref{eq_integral_capi}), so that we can use Eq.~(\ref{wintegrals2}).
    We start by writing the relative two-boson Green's function in terms of an infinite sum,
    \begin{eqnarray}
    \label{eq_greens_sum}
        G^{(\text{2B})}\lp E-E_{\ell},0,~\frac{\sqrt{3}}{2}u\rp
        = \sum_{p=0,2,\dots} \frac{\phi_p^*\lp\frac{\sqrt{3}}{2}u\rp\phi_p\lp 0\rp}{E_p-(E-E_{\ell})} . \nonumber \\
        \end{eqnarray}
    Inserting Eq.~(\ref{eq_greens_sum}) into
Eq.~(\ref{eq_integral_capi}) and writing the harmonic oscillator functions in terms of Hermite polynomials, we have
\begin{widetext}
            \begin{align}
                I_{\ell k}^{\text{(3B)}}(E-E_{\ell}) &=\sum_{p=0,2,\dots} \frac{\phi_p\lp 0\rp A_\ell A_k A_p}{E_p-(E-E_{\ell})} \int_{-\infty}^\infty H_\ell\lp -\frac{u}{2 a_{\text{ho}}}\rp H_k\lp \frac{u}{a_{\text{ho}}}\rp H_p\lp \frac{\sqrt{3}u}{2a_{\text{ho}}}\rp \exp \left[-\left(\frac{u}{a_{\text{ho}}}\right)^2\right] du\label{tmp1},
                \end{align}
where $A_k$ denotes a  dimensionful normalization constant,
\begin{eqnarray}A_k=\sqrt{\frac{1}{a_{\text{ho}}\sqrt{\pi}2^{k}k!}}.
\end{eqnarray}
Making the variable substitution $u=\sqrt{2}u'$ and subsequently setting $u'=u$, Eq. \eqref{tmp1} becomes
            \begin{align}\label{tmp2}
                I_{\ell k}^{\text{(3B)}}(E-E_{\ell}) &= \sum_{p=0,2,\dots} \frac{\sqrt{2}\phi_p\lp 0\rp A_\ell A_k A_p}{E_p-(E-E_{\ell})} \int_{-\infty}^\infty H_\ell\lp -\frac{u}{\sqrt{2}a_{\text{ho}}}\rp H_k\lp \frac{\sqrt{2}u}{a_{\text{ho}}}\rp H_p\lp \frac{\sqrt{6}u}{2a_{\text{ho}}}\rp \exp \left[-2\left( \frac{u}{a_{\text{ho}}}\right)^2\right] du.
            \end{align}
\end{widetext}
To be able to apply  Eq.~(\ref{wintegrals2}), we need to modify the arguments of the Hermite polynomials. This can be accomplished using~\cite{NIST:DLMF} 
\begin{align}\label{hermiterelation}
            H_k(\alpha x) =\sum_{v_k=0}^{\flr{\frac{k}{2}}} \gamma_{v_k}^{(k)}(\alpha) H_{k-2v_k}(x),
        \end{align}
    where $\alpha$ is a real number, $v_k$ takes the values $0,1,\cdots$, and $\gamma_{v_k}^{(k)}(\alpha)$ is defined as  
        \begin{align}
            \gamma_{v_k}^{(k)}(\alpha) = \alpha^{k-2v_k}\lp \alpha^2-1\rp^{v_k} \binom{k}{2v_k} \frac{(2v_k)!}{v_k!}
        \end{align}
    (we use the definition $0^0=1$). In Eq.~(\ref{hermiterelation}), the function \(\flr{\cdot}\) denotes the greatest integer  or floor function. Given a real number $x$, \(\flr{x}\) returns the greatest integer less than or equal to \(x\) (e.g., \(\flr{2}=2\) and \(\flr{\pi}=3\)).
    Applying Eq.~(\ref{hermiterelation}) to the three Hermite polynomials in the integrand of Eq.~(\ref{tmp2}), we obtain
\begin{widetext}    \begin{align}\label{bbbrecursiionrewrite}
            I_{\ell k}^{\text{(3B)}}(E-E_{\ell})=\sum_{p=0,2,\dots} \frac{\sqrt{2}\phi_p\lp 0\rp A_\ell A_k A_p}{E_p-(E-E_{\ell})} \sum_{v_{\ell}=0}^{\flr{\frac{\ell}{2}}}\sum_{v_{k}=0}^{\flr{\frac{k}{2}}}\sum_{v_{p}=0}^{\flr{\frac{p}{2}}}\gamma_{v_\ell}^{(\ell)}\lp -\frac{1}{\sqrt{2}}\rp\gamma_{v_k}^{(k)}\lp \sqrt{2}\rp \gamma_{v_p}^{(p)}\lp \frac{\sqrt{6}}{2}\rp \nonumber \\\times\int_{-\infty}^\infty H_{\ell-2v_\ell}\lp\frac{u}{a_{\text{ho}}}\rp H_{k-2v_k}\lp \frac{u}{a_{\text{ho}}}\rp H_{p-2v_p}\lp\frac{u}{a_{\text{ho}}}\rp \exp\lb -2\left(\frac{u}{a_{\text{ho}}}\right)^2\rb du.
            \end{align}
           Using Eq.~(\ref{wintegrals2}), our final expression for $I_{\ell k}^{\text{(3B)}}(E-E_{\ell})$ reads  
           \begin{align}
           \label{bbbrecursiionrewrite2}
            I_{\ell k}^{\text{(3B)}}(E-E_{\ell})=\nonumber \\
            \sum_{p=0,2,\dots} \frac{\sqrt{2}a_{\text{ho}} A_\ell A_k A_p\phi_p\lp 0\rp}{E_p-(E-E_{\ell})} \sum_{v_{\ell}=0}^{\flr{\frac{\ell}{2}}}\sum_{v_{k}=0}^{\flr{\frac{k}{2}}}\sum_{v_{p}=0}^{\flr{\frac{p}{2}}}\gamma_{v_\ell}^{(\ell)}\lp -\frac{1}{\sqrt{2}}\rp\gamma_{v_k}^{(k)}\lp \sqrt{2}\rp \gamma_{v_p}^{(p)}\lp \frac{\sqrt{6}}{2}\rp \times \frac{W_{\ell-2v_\ell,k-2v_k,p-2v_p,0}}{A_{\ell-2v_\ell,k-2v_k,p-2v_p,0}}.
        \end{align}\end{widetext}
        The key accomplishment of our manipulations is that Eq.~(\ref{bbbrecursiionrewrite2}) expresses the integral $I^{(\text{3B})}_{\ell k}(E-E_{\ell})$ in terms of $W_{\bullet}$, which can---as discussed in the next section---be determined efficiently.

The structure of the integrals $J^{\text{(3F)}}_{\ell k}(E-E_{\ell})$ and $K^{\text{(3F)}}_{\ell k}(E-E_{\ell})$, Eqs.~(\ref{eq_integral_capj}) and (\ref{eq_integral_capk}), is similar to that of  $I^{\text{(3B)}}_{\ell k}(E-E_{\ell})$, with two main differences:  a derivative operator is applied to one of the harmonic oscillator states as well as to the fermionic Green's function. To remove the derivative of the harmonic oscillator function, we use
 \begin{align}\label{derphi}
            \frac{\partial \phi_k(x)}{\partial x}= A_k \exp \left[-\frac{1}{2}\left(\frac{x}{a_{\text{ho}}}\right)^2\right] \times \nonumber \\
            \lb \frac{2k}{a_{\text{ho}}}H_{k-1}\lp\frac{x}{a_{\text{ho}}}\rp -\frac{x}{a_{\text{ho}}^2}H_k\lp \frac{x}{a_{\text{ho}}}\rp\rb.
        \end{align}
        The derivative of the Green's function can be written as an infinite series, 
        \begin{eqnarray}
        \label{eq_green_odd_series}
            \left[ \frac{\partial G^{(\text{2F})}\lp E-E_{\ell},z,\frac{\sqrt{3}}{2}u\rp}{\partial z} \right]_{z=0}=
            \nonumber \\
            \sum_{p=1,3,\dots}\frac{\phi_{p}^*\lp \frac{\sqrt{3}}{2}u\rp \left[\frac{\partial \phi_p(z)}{\partial z}\right]_{z=0}}{E_p-(E-E_{\ell})}.
        \end{eqnarray}
        Equation~(\ref{derphi}) shows that the integrals $J^{\text{(3F)}}_{\ell k}(E-E_{\ell})$ and $K^{\text{(3F)}}_{\ell k}(E-E_{\ell})$ that enter into  Eq.~\eqref{goaleqfff} can each be rewritten as two integrals. This gives us four integrals altogether that enter into the FFF eigenequation. Each of these integrals can be computed following steps similar to those employed above when evaluating the integral  $I^{\text{(3B)}}_{\ell k}(E-E_{\ell})$ for the BBB system. One notable modification originates in  the fact that the second Hermite polynomial on the right-hand side of Eq.~(\ref{derphi}) is multiplied by $x$. To deal with this factor,  we recognize that \(x\) can be expressed as \(\frac{a_{\text{ho}}}{2}H_{1}\lp \frac{x}{a_{\text{ho}}}\rp\). The introduction of another Hermite polynomial does not introduce any additional challenges, since $W_{\bullet}$, Eq.~(\ref{wintegrals2}), is defined as an integral over four Hermite polynomials.
        Using Eqs.~(\ref{derphi}) and (\ref{eq_green_odd_series}) in Eq.~(\ref{goaleqfff}) and following steps similar to those for evaluating  the BBB integral $I^{\text{(3B)}}_{\ell k}(E-E_{\ell})$, we find
\begin{widetext}
    \begin{eqnarray}\label{JIntexpansion}
        J^{\text{(3F)}}_{\ell k}(E-E_{\ell})= \nonumber \\
        \sum_{p=1,3,5,\dots} \frac{\left[ \frac{\partial \phi_p(z)}{\partial z} \right]_{z=0} A_p A_{\ell} A_k}{E_p-(E-E_{\ell})} \Bigg[ 2\sqrt{2} k \sum_{v_k=0}^{\flr{\frac{k-1}{2}}}  \sum_{v_\ell=0}^{\flr{\frac{\ell}{2}}} \sum_{v_p=0}^{\flr{\frac{p}{2}}} \gamma_{v_k}^{(k-1)}\lp \sqrt{2}\rp  \gamma_{v_\ell}^{(\ell)}\lp -\frac{1}{\sqrt{2}}\rp   \gamma_{v_p}^{(p)} \lp \frac{\sqrt{6}}{2}\rp \times\frac{W_{k-1-2v_k,\ell-2v_\ell,p-2v_p,0}}{A_{k-1-2v_k,\ell-2v_\ell,p-2v_p,0}} \nonumber  \\
        -\sum_{v_k=0}^{\flr{\frac{k}{2}}}  \sum_{v_\ell=0}^{\flr{\frac{\ell}{2}}} \sum_{v_p=0}^{\flr{\frac{p}{2}}} \gamma_{v_k}^{(k)}\lp \sqrt{2}\rp  \gamma_{v_\ell}^{(\ell)}\lp -\frac{1}{\sqrt{2}}\rp   \gamma_{v_p}^{(p)} \lp \frac{\sqrt{6}}{2}\rp \times \frac{W_{k-2v_k,\ell-2v_\ell,p-2v_p,1}}{A_{k-2v_k,\ell-2v_\ell,p-2v_p,1}}\Bigg] \nonumber \\
    \end{eqnarray}
and
        \begin{eqnarray}\label{KIntexpansion}
             K^{\text{(3F)}}_{\ell k}(E-E_{\ell})=  \nonumber \\
             \sum_{p=1,3,5,\dots} \frac{\left[\frac{\partial \phi_p(z)}{\partial z} \right]_{z=0} A_p A_{\ell} A_k}{E_p-(E-E_{\ell})} \Bigg[ 2\sqrt{2} \ell \sum_{v_k=0}^{\flr{\frac{k}{2}}}  \sum_{v_\ell=0}^{\flr{\frac{\ell-1}{2}}} \sum_{v_p=0}^{\flr{\frac{p}{2}}} \gamma_{v_k}^{(k)}\lp \sqrt{2}\rp  \gamma_{v_\ell}^{(\ell-1)}\lp -\frac{1}{\sqrt{2}}\rp   \gamma_{v_p}^{(p)} \lp \frac{\sqrt{6}}{2}\rp \times\frac{W_{k-2v_k,\ell-1-2v_\ell,p-2v_p,0}}{A_{k-2v_k,\ell-1-2v_\ell,p-2v_p,0}} 
             \nonumber \\
        +\frac{1}{2}\sum_{v_k=0}^{\flr{\frac{k}{2}}}  \sum_{v_\ell=0}^{\flr{\frac{\ell}{2}}} \sum_{v_p=0}^{\flr{\frac{p}{2}}} \gamma_{v_k}^{(k)}\lp \sqrt{2}\rp  \gamma_{v_\ell}^{(\ell)}\lp -\frac{1}{\sqrt{2}}\rp   \gamma_{v_p}^{(p)} \lp \frac{\sqrt{6}}{2}\rp \times \frac{W_{k-2v_k,\ell-2v_\ell,p-2v_p,1}}{A_{k-2v_k,\ell-2v_\ell,p-2v_p,1}}\Bigg]. \nonumber \\
        \end{eqnarray}
\end{widetext}
     The next section discusses implementation details and  the convergence behavior of the matrix elements $A_{{\ell} k}^{(\text{3F})}(E-E_{\ell})$ with increasing $p_{\text{max}}$, where $p_{\text{max}}$ is the largest $p$ value included in Eqs.~(\ref{JIntexpansion}) and (\ref{KIntexpansion}).

To evaluate the integral $I_{stk \ell}^{(\text{4B})}(E-E_s-E_t)$, Eq.~(\ref{eq_integral_capi4}), we represent the relative two-boson Green's function $G^{(\text{2B})}(E-E_s-E_t,0,(u+\sqrt{2}v)/2)$ using the infinite series representation given in Eq.~(\ref{eq_greens_sum}). 
Importantly, the exponential functions contained in the integrand on the right-hand side of Eq.~(\ref{eq_integral_capi4}), i.e., those contained in the  Green's function and the four harmonic oscillator functions, combine to 
$\exp \left[-(u^2+v^2)/a_{\text{ho}}^2 \right]$. 
 This indicates that the exponential term can be written as a product of an exponential in $u$ and an exponential in $v$. The integrand does, however, contain two Hermite polynomials that depend on both $u$ and $v$. To ``isolate'' the $u$ and $v$ dependencies, we use~\cite{A-S}
\begin{eqnarray}
    H_k\left(\frac{u \pm \sqrt{2}v}{2}\right) = \nonumber \\
    2^{-k/2}\sum_{k'=0}^k \binom{k}{k'}H_{k'}\lp \frac{u}{\sqrt{2}}\rp H_{k-k'}\lp \pm v\rp .
\end{eqnarray}
As a result, we can rewrite the two-dimensional integral over $u$ and $v$ on the right-hand side of Eq.~(\ref{eq_integral_capi4}) as a product of a one-dimensional integral over $u$ and a one-dimensional integral over $v$. Collecting terms, we find

        \begin{widetext}
        \begin{align}\label{onetotwo}
            I_{stk \ell}^{(\text{4B})}(E-E_s-E_t)=
            \sum_{p=0,2,\dots} \frac{\phi_p(0) A_p A_{s} A_t A_k A_\ell}{E_p-(E-E_s-E_t)} 2^{-\lp p+s\rp/2}\sum_{\gamma_p=0}^p \sum_{\gamma_s=0}^s \binom{p}{\gamma_p}\binom{s}{\gamma_s} \times \nonumber \\ \int_{-\infty}^\infty H_{\gamma_p}\lp \frac{v}{a_{\text{ho}}}\rp H_{\gamma_s}\lp -\frac{v}{a_{\text{ho}}}\rp H_{\ell}\lp \frac{v}{a_{\text{ho}}}\rp \exp\left[-\left(\frac{v}{a_{\text{ho}}} \right)^2 \right]d v \nonumber\\
            \times \int_{-\infty}^\infty H_{p-\gamma_p}\lp \frac{u}{\sqrt{2} a_{\text{ho}}}\rp H_{s-\gamma_s}\lp \frac{u}{\sqrt{2} a_{\text{ho}}}\rp H_{t}\lp -\frac{u}{\sqrt{2}a_{\text{ho}}}\rp H_{k}\lp \frac{u}{a_{\text{ho}}} \rp \exp \left[-\left(\frac{u}{a_{\text{ho}}} \right)^2 \right] d u,
        \end{align}
        where $\gamma_p$ and $\gamma_s$ increase in steps of $1$.
The one-dimensional integrals in Eq.~(\ref{onetotwo}) are of the form 
    \begin{align}
        C_{i,j,k,\ell,\alpha,\beta,\gamma,\delta,\varepsilon}=\intinf H_{i}\lp \frac{\alpha u}{a_{\text{ho}}}\rp H_{j}\lp \frac{\beta u}{a_{\text{ho}}}\rp H_{k}\lp \frac{\gamma u}{a_{\text{ho}}}\rp H_{\ell}\lp \frac{\delta u}{a_{\text{ho}}}\rp \exp\lp -\varepsilon\frac{u^2}{a_{\text{ho}}^2}\rp~du.
    \end{align}
Using the techniques introduced above, we find
    \begin{align}
        \label{eq_general_int_expression}
        C_{i,j,k,\ell,\alpha,\beta,\gamma,\delta,\varepsilon}= \nonumber \\
        a_{\text{ho}}\sqrt{\frac{2}{\varepsilon}}\sum_{i'=0}^{\flr{\frac{i}{2}}} \sum_{j'=0}^{\flr{\frac{j}{2}}} \sum_{k'=0}^{\flr{\frac{k}{2}}} \sum_{\ell'=0}^{\flr{\frac{\ell}{2}}} \gamma_{i'}^{(i)}\lp \alpha\sqrt{\frac{2}{\varepsilon}}\rp  \gamma_{j'}^{(j)}\lp \beta\sqrt{\frac{2}{\varepsilon}}\rp   \gamma_{k'}^{(k)}\lp \gamma\sqrt{\frac{2}{\varepsilon}}\rp   \gamma_{\ell'}^{(\ell)}\lp \delta\sqrt{\frac{2}{\varepsilon}}\rp  \times \frac{W_{i-2i',j-2j',k-2k',\ell-2\ell'}}{A_{i-2i',j-2j',k-2k',\ell-2\ell'}}.
    \end{align}
\end{widetext}
    Using Eq.~(\ref{eq_general_int_expression}) twice in Eq.~(\ref{onetotwo}) and taking advantage of the fact that the Hermite polynomials are, depending on their order, even or odd functions, yields an expression for $I_{stk \ell}^{(\text{4B})}(E-E_s-E_t)$ in terms of one infinite sum [the sum over $p$ in Eq.~(\ref{onetotwo})] and six finite sums [the sums over $\gamma_p$ and $\gamma_s$ in Eq.~(\ref{onetotwo}); three sums that arise from changing the 
    argument of the first exponential in Eq.~(\ref{onetotwo}) to $-2(v/a_{\text{ho}})^2$; and one sum that arises from changing the 
    argument of the second exponential in Eq.~(\ref{onetotwo}) to $-2(u/a_{\text{ho}})^2$].  
     The next section discusses implementation details.

\subsection{Numerical considerations and convergence analysis}

The scheme for determining the eigenenergies and eigenstates of the BBB, FFF, and BBBB systems---which was developed in the previous sections---depends on two cutoffs, namely the size of the basis set [dimension of the square matrices with elements
$A_{\ell k}^{(\text{3B})}(E-E_{\ell})$, $A_{\ell k}^{(\text{3F})}(E-E_{\ell})$, and $B_{\ell k}^{(\text{4B})}(E-E_s-E_t)$; see Eqs.~(\ref{goaleqbbb}), (\ref{goaleqfff}), and (\ref{goaleqbbbb})] and the largest $p$ value, denoted by $p_{\text{max}}$, that is used when calculating $I_{\ell k}^{(\text{3B})}(E-E_{\ell})$, $J_{\ell k}^{(\text{3F})}(E-E_{\ell})$, $K_{\ell k}^{(\text{3F})}(E-E_{\ell})$, and $I_{\ell k}^{(\text{4B})}(E-E_s-E_t)$ [see Eqs.~(\ref{bbbrecursiionrewrite2}), (\ref{JIntexpansion}), (\ref{KIntexpansion}), and (\ref{onetotwo})]. The results in Sec.~\ref{sec_spectra} are obtained using C++. We checked our C++ implementations for the BBB, FFF, and BBBB systems  against Python and Mathematica implementations, where the latter was run in a mode that ensures many more digits of accuracy than the C++ implementation.  

    A salient feature of Eqs. \eqref{bbbrecursiionrewrite2}, (\ref{JIntexpansion}), and (\ref{KIntexpansion}) is that the   sums over $v_\ell$, $v_k$, and $v_p$ are independent of the relative energy $E$. Analogously, all sums in Eq.~(\ref{onetotwo}), except for the sum over $p$, are independent of the relative energy $E$. This allows us to precalculate and store the energy-independent terms and recycle them  for different  \(E\). The recursive evaluation of the \(W_\bullet\) factors [see Eq.~(\ref{wintegralsrecursion2})], where ``$\bullet$''  collectively denotes the four subscripts,   is  computationally  very efficient. Our C++ implementation stores all auxiliary quantities, including integers, as {\em{doubles}}, which occupy 64 bits of memory, or as ``doubles'' with ``boosted precision'' (see below).  
    
    The two cutoffs (basis set size and $p_{\text{max}}$) are not independent since an accurate determination of a matrix element of ``larger order'' (e.g., larger value of $\ell+k$) requires a larger $p_{\text{max}}$ than the determination of a matrix element of ``lower order'' (e.g., smaller value of $\ell+k$). For the BBB and BBBB systems, our C++ implementation that utilizes the datatype {\em{doubles}} consistently yields  numerically stable and converged results for energies $E$ between $-10$~$\hbar \omega$ and $15$~$\hbar \omega$ and $g_+$ between $-10$~$\hbar \omega a_{\text{ho}}$ and $10$~$\hbar \omega a_{\text{ho}}$. For the FFF system, obtaining numerically stable and converged results is, as we now illustrate,  more challenging.

\begin{figure}
    \centering
    \includegraphics[width=\linewidth]{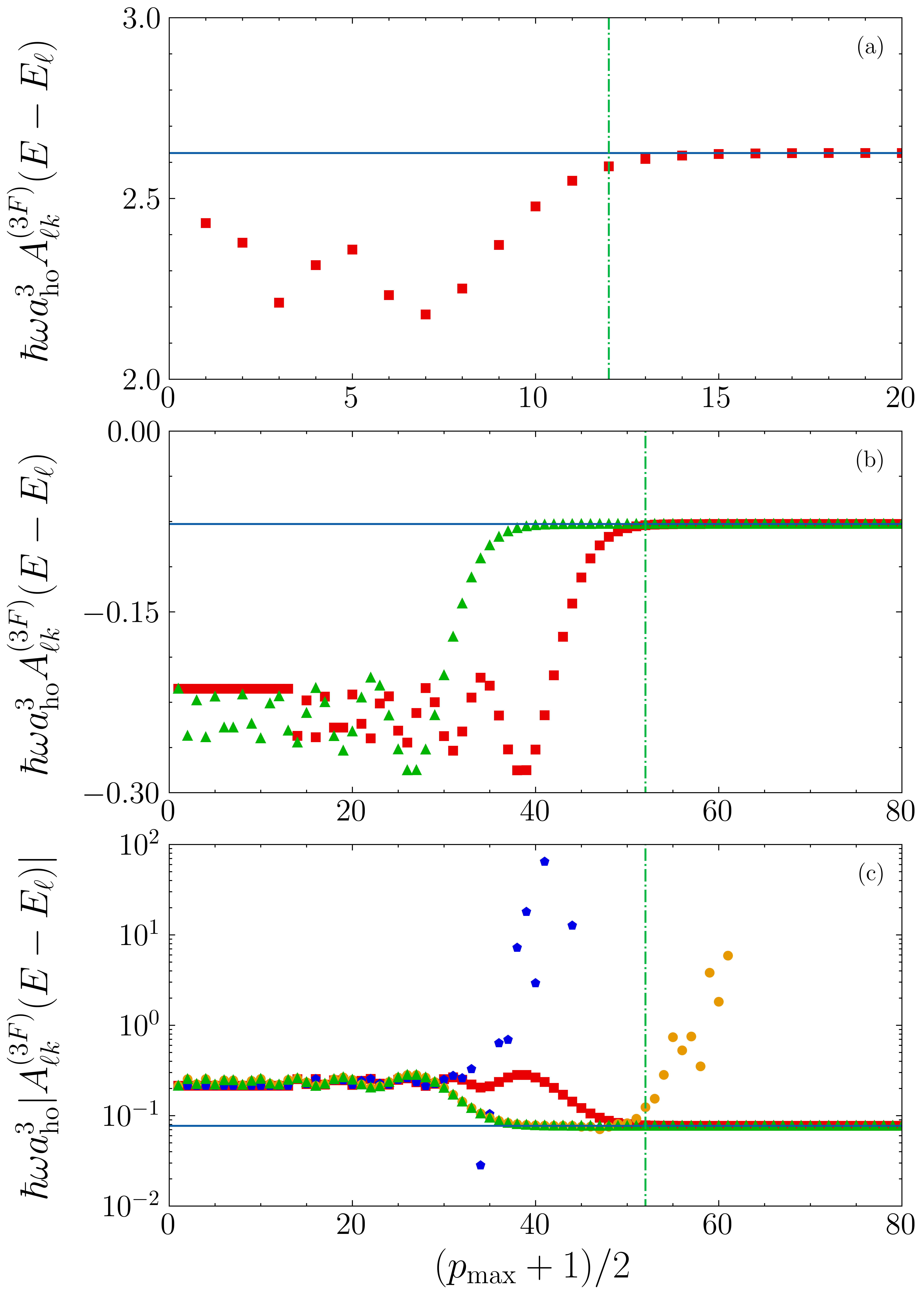}    
    \caption{Convergence analysis of the matrix elements $A_{\ell k}^{(\text{3F})}(E-E_\ell)$ for $E=2.5001$~$\hbar \omega$ as a function of $(p_{\text{max}}+1)/2$ for the FFF system; since the sum over $p$ in Eqs.~(\ref{JIntexpansion}) and (\ref{KIntexpansion}) runs over odd $p$, the quantity $(p_{\text{max}}+1)/2$ corresponds to the number of terms included in the evaluation of the integrals $J_{\ell k}^{(\text{3F})}(E-E_{\ell})$ and $K_{\ell k}^{(\text{3F})}(E-E_{\ell})$. Symbols are obtained using our C++ implementation that is based on the  recursion relationship introduced in this work. For comparison, the horizontal blue solid   lines show  the corresponding fully converged matrix elements, calculated using Mathematica. The vertical green dashed  lines mark the heuristic convergence estimate $p_{\text{max}}+1=2(k+\ell)$; our FFF calculations use a $p_{\text{max}}$ such that $p_{\text{max}}+1$ is larger than $2(k+\ell)$.
    (a)  Red squares show $A^{(\text{3F})}_{\ell k}(E-E_{\ell})$ for  $k=\ell=6$, calculated using {\em{Boost Multiprecision}} library functions that work with 50 decimal digits of precision. (b) Green triangles  show $A^{(\text{3F})}_{\ell k}(E-E_{\ell})$ for  $\ell=38$ and $k=14$ 
    while red squares 
    show $A^{(\text{3F})}_{\ell k}(E-E_{\ell})$ for $\ell=14$ and $k=38$, calculated using {\em{Boost Multiprecision}} library functions that work with 50 decimal digits of precision. (c)
    The green triangles and red squares show the magnitude of $A^{(\text{3F})}_{\ell k}(E-E_{\ell})$ [same data as in panel (b)], using a log-scale for the vertical axis. For comparison, blue pentagons and orange circles  show the same matrix elements but calculated using the datatype {\em{double}}. It can be seen that the datatype {\em{double}} is plagued by numerical instabilities for large $(p_{\text{max}}+1)/2$: the matrix elements diverge for $(p_{\text{max}}+1)/2 \gtrsim 30$ for $\ell=14$ and $k=38$ and  for $(p_{\text{max}}+1)/2 \gtrsim 45$ for $\ell=38$ and $k=14$.        }
    \label{fig:placeholder}
\end{figure}

    Due to the discontinuities of some of the functions that enter into the FFF eigenvalue equation, the outlined approach for calculating the matrix elements $A_{\ell k}^{(\text{3F})}(E-E_{\ell})$ for ``large'' values of $\ell + k$ is numerically less stable than that for calculating the  matrix elements $A_{\ell k}^{(\text{3B})}(E-E_{\ell})$ for the same $\ell$ and $k$. To eliminate numerical instabilities, our three-fermion C++ implementation utilizes {\em{Boost Multiprecision}}  library functions that work with 50 decimal digits of precision. 
To illustrate the convergence of the FFF matrix elements with increasing cutoff $p_{\text{max}}$, red dots in Fig.~\ref{fig:placeholder}(a) show the matrix element $A_{6,6}^{(\text{3F})}(E-E_6)$ for  $E=2.5001$~$\hbar \omega$  as a function of $(p_{\text{max}}+1)/2$. It can be seen that the red dots approach the converged results (horizontal blue solid line) with increasing   $(p_{\text{max}}+1)/2$ in an oscillatory manner. Rough convergence is obtained for $(p_{\text{max}}+1)/2$ about $12$, which is shown by the dashed vertical line [this can be expressed as  $(p_{\text{max}}+1)/2=\ell+k$]. We found that this estimate of when rough convergence is reached holds quite generally also for other  $A_{\ell k}^{(\text{3F})}(E-E_{\ell})$. Correspondingly, we typically set  $p_{\text{max}}+1$ to be larger than $2(\ell+k)$. For  $(\ell,k)=(6,6)$ [Fig.~\ref{fig:placeholder}(a)], the datatype {\em{doubles}} yields the same results as our increased precision implementation that uses {\em{Boost Multiprecision}} library functions. 
Red circles and green squares in Fig.~\ref{fig:placeholder}(b) 
show the matrix elements $A_{\ell k}^{(\text{3F})}(E-E_{\ell})$ for $(\ell,k)=(38,14)$ and $(\ell,k)=(14,38)$. These  results are obtained using {\em{Boost Multiprecision}} library functions. It can be seen that both elements converge to the same value (horizontal blue solid line), as required for a well-defined Hamiltonian. Interestingly, the convergence behavior of the two matrix elements differs. Utilizing instead the datatype {\em{doubles}}, leads to numerical instabilities. Blue pentagons and orange circles in Fig.~\ref{fig:placeholder}(c) show that the sums, calculated using the datatype {\em{doubles}}, diverge for $(p_{\text{max}}+1)/2 \gtrsim 30$ and $50$, respectively. For the blue pentagons, the divergence occurs prior to the matrix element reaching convergence. For the orange circles, instead, the divergence occurs after the matrix element reaches convergence. 
    The fact that the computationally less expensive {\em{doubles}} implementation yields converged results for $A_{\ell k }^{(\text{3F})}(E-E_{\ell})$ for $\ell > k$ suggests that the FFF spectrum can be determined using {\em{doubles}} if $p_{\text{max}}+1$ is chosen  carefully and if 
    $A_{\ell k }^{(\text{3F})}(E-E_{\ell})$ for 
    $k > \ell$
    is replaced by the corresponding converged result for $\ell > k$.

\section{Energy Spectra}
\label{sec_spectra}
The approach discussed in the previous section allows for an efficient determination of the BBB ($\Pi=+1$), FFF ($\Pi=-1$), and BBBB ($\Pi=+1$) energy spectra and eigenstates. This section shows results only for the energy spectra and not for the eigenstates. 
While the BBB energy spectrum has been reported in the literature~\cite{10.1088/0953-4075/47/6/065303}, we are not aware of work that shows the entire energy spectrum of the BBBB system (up to a given energy cutoff) nor are we aware of any work that  calculates the FFF energy spectrum directly, i.e., without employing the celebrated Bose-Fermi mapping; we note, though, that variational approaches exist~\cite{10.1088/1367-2630/abb386}.  By  calculating the FFF spectrum directly, we aim to show that the odd-parity pseudopotential does not, if treated carefully, introduce any conceptual issues within the Lippmann-Schwinger equation framework. This opens the door for applying the approaches developed in our work to  1D systems with even- and odd-parity interactions and for guiding work that is aimed at treating single-species fermions using the framework of second quantization. The Bose-Fermi mapping, which allows one to determine the FFF spectrum from the BBB spectrum, is utilized below to establish the correctness of our FFF results.   

\begin{figure}[hbt!]
        \centering
       \includegraphics[width=\linewidth]{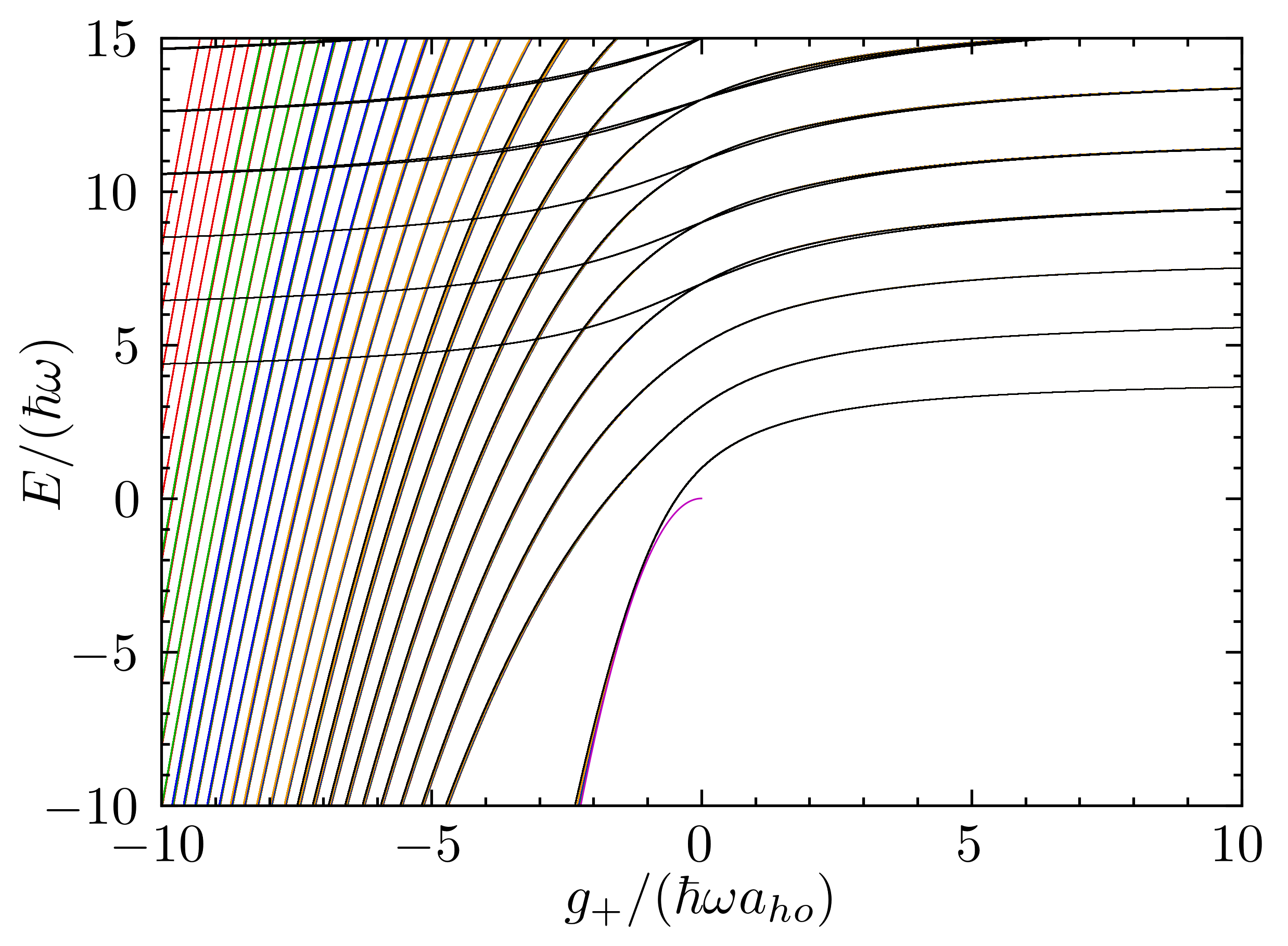}
        \caption{Energy spectra of the BBB system with positive relative parity ($\Pi=+1$) as a function of the coupling constant $g_+/(\hbar \omega a_{\text{ho}})$ for five different basis set sizes. The black, brown, blue, green, and red dots show  spectra for $\ell_{\text{max}}=k_{\text{max}}=20$, $30$,  $40$, $50$, and $60$, respectively (in each case,  $\ell_{\text{max}}/2$ eigenenergies are shown). The spectra are calculated using $p_{\text{max}}=2(\ell_{\text{max}}+k_{\text{max}})$. $E$ denotes, consistent with the notation in the text, the relative energy; the center-of-mass energy is not included in the spectra. The energy spectra for the five basis set sizes agree  very well over the energy range shown. In fact, the agreement is so good, that the energies of the low-lying states for $\ell_{\text{max}}=k_{\text{max}}=60$, e.g.,  are ``covered'' by those calculated for smaller basis set sizes. 
         The slight spread of the energies of the lowest state (``thickening of the line'') for $E$ around $-5$ to $-10$~$\hbar \omega$ 
         indicates that convergence in this regime requires $\ell_{\text{max}}=k_{\text{max}} \gtrsim 30$. For comparison, the purple solid line on the negative $g_+$ side shows the binding energy of three identical bosons in free space.
        }
        \label{3Bcon}
    \end{figure}

\begin{figure}[hbt!]
    \centering
    \includegraphics[width=\linewidth]
    {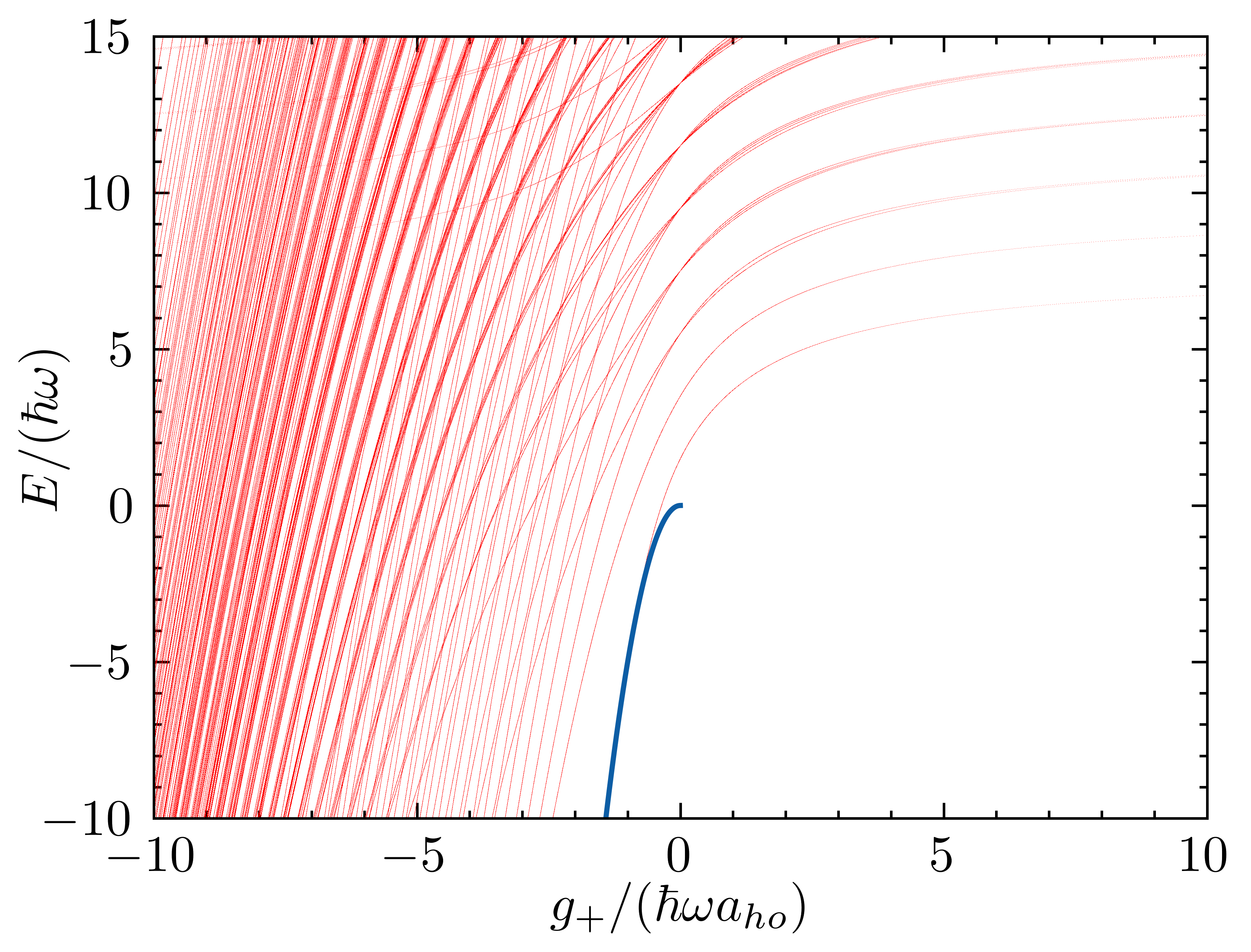}
    \caption{Energy spectrum of the BBBB system with positive relative parity ($\Pi=+1$) as a function of the coupling constant $g_+/(\hbar \omega a_{\text{ho}})$ (red dots). The spectrum is calculated using $s_{\text{max}}=t_{\text{max}}=k_{\text{max}}=\ell_{\text{max}}=32$ and $p_{\text{max}}=192$.
    For comparison, the blue line shows the binding energy of the BBBB system in free space. It can be seen that the tetramer-like state of the trapped system agrees very well with the binding   energy of the free-space tetramer. Since our approach only explicitly builds in two-body correlations from the outset and not three- or four-body correlations, the excellent agreement on the negative $g_+$-side provides  non-trivial confirmation that our basis effectively  captures all system correlations. 
    }
    \label{4B}
\end{figure}

    In addition, we performed the following checks. (i) We checked that  the degeneracies of the $g_{\pm}=0$ and  $|g_{\pm}|=\infty$ spectra agree with those determined analytically. (ii) We compared several energetically  low-lying energy levels for small $|g_{\pm}|$ and small $|g_{\pm}|^{-1}$ with those determined within first-order time-independent degenerate perturbation theory. Specifically, we confirmed that the energy levels are varying, to a good approximation, linearly with $g_{\pm}$ for small $|g_{\pm}|$ and also linearly with $(g_{\pm})^{-1}$ for small $|g_{\pm}|^{-1}$, in agreement with the slopes determined perturbatively (for this analysis, we employed techniques similar to those discussed in Refs.~\cite{https://doi.org/10.1038/ncomms6300, PhysRevA.91.013620}).
    (iii) We adapted our numerical approach to the eigenvalue equations, derived in Ref.~\cite{PhysRevLett.111.045302}, for the two-component FFF' and FFF'F' systems, in which unlike fermions F and F' interact through the 1D even-parity pseudopotential. Excellent agreement was found with the energy spectra determined by evaluating the integrals that enter into the eigenvalue equations by alternative, computationally less efficient techniques~\cite{PhysRevLett.111.045302, PhysRevA.86.042702}.
    (iv) We checked that the lowest energy state approaches the binding energy of the three- and four-body bound states in free space~\cite{https://doi.org/10.1063/1.1704156} (the three- and four-boson comparisons are shown in Figs.~\ref{3Bcon} and \ref{4B}, respectively). 

The determination of the three-boson spectrum involves two infinite sums: the sum over $k$ in Eq.~(\ref{goaleqbbb}) and the sum over $p$ in Eq.~(\ref{bbbrecursiionrewrite}).  Figure~\ref{3Bcon} shows
     the BBB energy spectrum 
     for five different basis set sizes, namely $k_{\text{max}}=20$ (black dots),  $30$ (brown dots), 
     $40$ (blue dots),
     $50$ (green dots),
     and $60$ (red dots). The spectra are calculated using $p_{\text{max}}=2(\ell_{\text{max}}+k_{\text{max}})$. We found that the spectra calculated using {\em{double}} and {\em{Boost precision}} data types are indistinguishable, indicating that the BBB matrix element determination and diagonalization are numerically much more robust than those for the FFF system. 
     The energy grid is chosen so densely (spacing of $0.0005$~$\hbar \omega$) that the individual dots cannot be seen in Fig.~\ref{3Bcon}. The merging of the dots into each other gives the appearance that the energy levels are represented by  lines. For positive coupling constant $g_+$, where the eigenstates all have gas-like characteristics (two- and three-body bound states are absent), the five basis set sizes yield results that are  indistinguishable by bare eye. Tiny differences can be detected for negative $g_+$, where all but one of the states that ``plunge down'' correspond to dimer plus atom states (for a similar analysis for three-dimensional systems, see Ref.~\cite{PhysRevA.81.053615}) and one state (namely, the one that has an energy of $E= \hbar \omega$ for $g_+=0$) corresponds to a bound trimer on the negative $g_+$-side. Looking at energies between $-10$ and $0$~$\hbar \omega$, we find that the energy of the bound trimer state  approaches that of three bound bosons in free space~\cite{https://doi.org/10.1063/1.1704156} (solid purple line). 

\begin{figure}[hbt!]
    \centering
    \includegraphics[width=\linewidth]{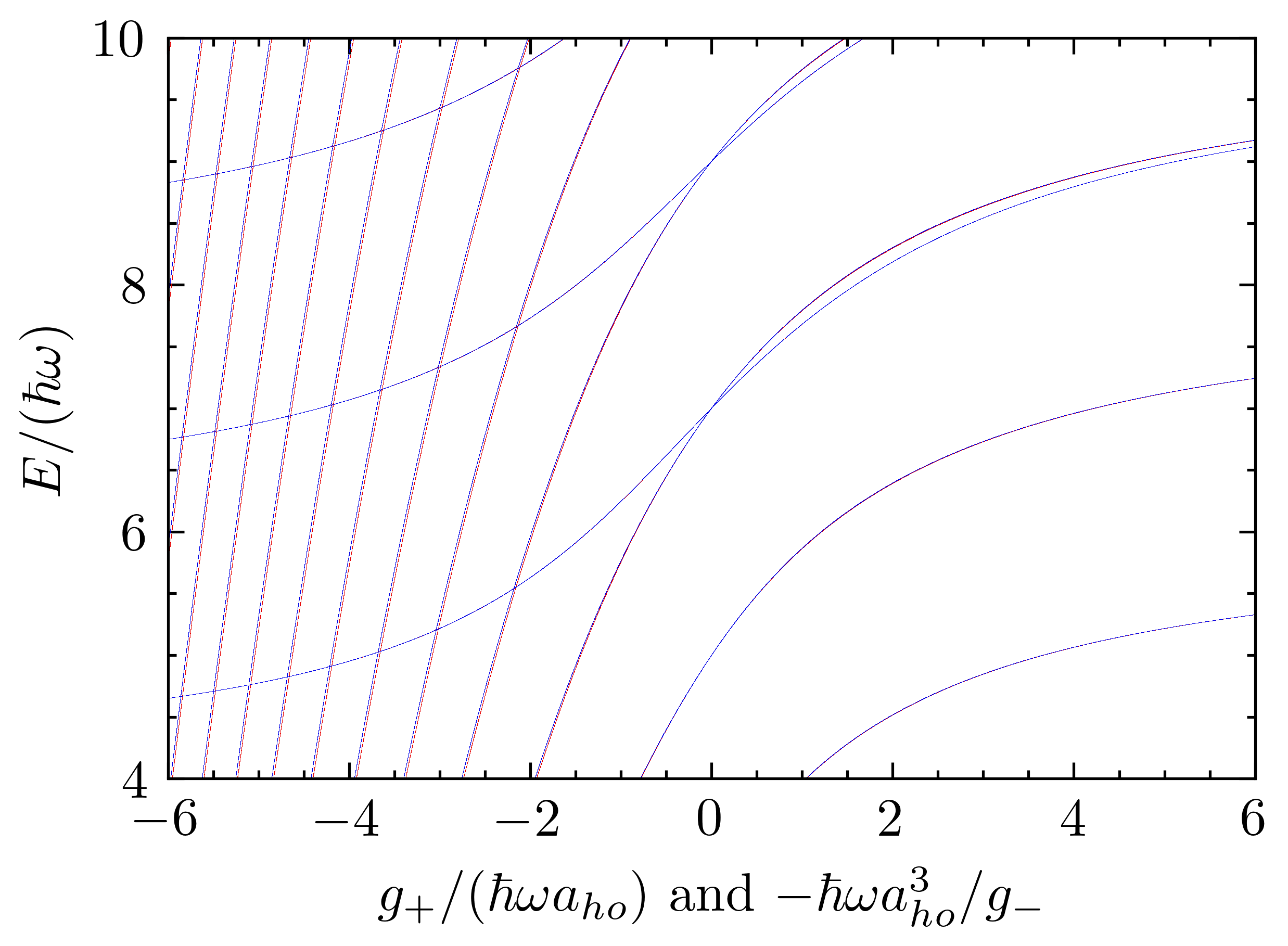}
    \caption{Energy spectra of the BBB system with positive relative parity ($\Pi=+1$) as a function of $g_+/(\hbar \omega a_{\text{ho}})$ (blue dots) and of the FFF system with negative relative parity ($\Pi=-1$) as a function of $-\hbar \omega a_{\text{ho}}^3/g_-$ (red dots). The spectra are calculated using $\ell_{\text{max}}=k_{\text{max}}=60$ and $p_{\text{max}}=240$. The C++ implementation uses {\em{Boost precision}}. As predicted by the Bose-Fermi mapping (see text), the  BBB and FFF spectra coincide in the chosen representation. The ``collapse'' of the spectra (the blue and red dots are essentially  indistinguishable on the scale shown) confirms the validity and accuracy of our calculations.  
    }
    \label{fbdual}
\end{figure}

     To benchmark the FFF energy spectrum, we use the Bose-Fermi mapping, which states that the energy spectra of identical 1D bosons and identical 1D fermions coincide, provided the 1D coupling constants are related to each other via
     $g_+ = - 2 \hbar^4 / (m^2 g_-)$~\cite{PhysRevA.70.023608, GIRARDEAU20043, PhysRevA.70.042709}. The relationship implies, e.g., that weakly-repulsive bosons behave like strongly-attractive fermions. Figure~\ref{fbdual} shows the BBB spectrum (blue dots) as a function of the dimensionless positive-parity coupling constant $g_+/(\hbar \omega a_{\text{ho}})$ together with the FFF spectrum (red dots), which is plotted as a function of the negative of the inverse of the dimensionless negative-parity coupling constant $-\hbar \omega a_{\text{ho}}^3/g_-$. Both spectra are calculated using the  {\em{Boost precision}} library functions, $\ell_{\text{max}}=k_{\text{max}}=60$, and $p_{\text{max}}=240$.  It can be seen that the energy spectra overlap essentially perfectly, thereby confirming the accuracy of these independently calculated spectra.

The red dots in Fig.~\ref{4B} show 
  the energy spectrum of the BBBB system (using an energy spacing of $0.005$~$\hbar \omega$) as a function of the coupling constant $g_+$.
  For positive $g_+$, the eigenstates correspond to gas-like  atom-atom-atom-atom states. For negative $g_+$, in contrast,
  the eigenstates can be categorized as follows (an analogous categorization of three-dimensional two-component Fermi gases can be found in Ref.~\cite{PhysRevA.81.053615}):
  (i) atom-atom-atom-atom states, (ii) atom-atom-dimer states, (iii) dimer-dimer states, (iv) atom-trimer states,    and (v) tetramer states. The energies of states with character~(i) are equal to $(15/2+2n)\hbar \omega$, where $n=0,1,\dots$, for $g_+=-\infty$.
  The energies of states with character (ii)-(v) approach $-\infty$ as $g_+$ approaches $-\infty$ since the energy of these states can be written in terms of that of bound dimers, trimers, and tetramers. The rate at which the energy of states with character (ii)-(v) ``dives down'' is different. Specifically, a close look at Fig.~\ref{4B} allows one to identify families of curves with different slopes. There exists exactly one state with character (v)~\cite{https://doi.org/10.1063/1.1704156}. This state has an energy of $3 \hbar \omega/2$ for $g_+=0$. On the negative $g_+$-side, the energy of this state agrees excellently with that of four bound bosons in free space (blue line in Fig.~\ref{4B}). While there exists exactly one bound trimer in free space for negative $g_+$, the addition of a fourth atom that is not bound to the trimer leads to a large number of states with character (iv); if the basis was infinitely large, the spectrum would contain an infinite number of states with atom-trimer character.  This family of states corresponds to the ``usual'' atom-trimer scattering continuum. In our case, the scattering continuum is discretized due to the presence of the harmonic confinement. Analogous arguments can be made for the atom-atom-dimer-like states [states with character~(ii)] and the dimer-dimer-like states [states with character~(iii)].

\section{Conclusions}
\label{sec_conclusion}
This work developed an efficient approach for solving the eigenvalue equations, derived using the Lippmann-Schwinger equation, for small single-component Bose and Fermi systems under external harmonic confinement. The scheme developed was applied to single-component systems that contain three or four particles. Since small effectively one-dimensional systems can be prepared in state-of-the-art cold-atom experiments~\cite{PhysRevLett.108.075303, PhysRevLett.94.210401, https://doi.org/10.1038/nature09036}, our work is relevant to both theory and experiment.

The single-component Bose and Fermi gas results can be utilized to treat anyons, utilizing the bosonic-anyon---fermionic-anyon mapping~\cite{hidalgosacoto2025universalmomentumtailidentical, hidalgosacoto2025identical1danyonszerorange}. The techniques developed might also prove useful in calculating the dynamics of few-body systems. Since the computational cost for calculating the four-particle energy spectrum is quite reasonable (calculations can be run on a laptop within a few minutes to a few hours, depending on the precision demanded)  extending the approach to the five-particle system seems feasible. The approach can also be extended to other parity sectors (e.g., negative parity for 1D bosonic systems) as well as one-dimensional mixtures that contain two types of bosons or two types of fermions, which can interact through even- and odd-parity pseudopotentials.
It would also be interesting to extend the techniques developed in this work to two- and three-dimensional systems as well as quasi-one- and quasi-two-dimensional systems.

\section{Acknowledgements}
We thank Tai Tran for sharing Ref.~\cite{son2024derivationrecursiveformulasintegrals} prior to publication. 
We gratefully acknowledge support by the National Science Foundation through grant number PHY-2409311.
We also acknowledge 
partial support by the National Science Foundation through grant number PHY-2243731 (REU/RET). 


\appendix
\onecolumngrid
\section{Derivation of FFF Eigenequation}
\label{appendixfff}
    This section derives the eigenequation for three identical fermions [see Eq.~\eqref{goaleqfff}]. 
    Since the odd-parity pseudopotential, Eq.~(\ref{vm}), depends on the derivative operators and since some of the functions involved have discontinuous derivatives, the derivation requires special attention.  One goal of the manipulations below is to recast integrands such that they are continuous functions in the variable that is being integrated over. Throughout this appendix, we employ the shorthand notations
        \begin{align}\label{shorthand}
            \partial_{i}f(x_1,\dots,x_i,\dots,x_n)=\lb \parder{y_i} f(y_1,\dots,y_i,\dots,y_n)\rb_{y_1=x_1,\dots,y_n=x_n}
        \end{align}
        and 
        \begin{align}\label{shorthand}
            \partial_{i,j}f(x_1,\dots,x_i,\dots,x_j,\dots,x_n)=\left\{\parder{y_j}\lb \parder{y_i} f(y_1,\dots,y_i,\dots,y_j,\dots,y_n)\rb \right\}_{y_1=x_1,\dots,y_n=x_n}
        \end{align}
    for multi-variable functions, where we implicitly assume $\partial y_i/\partial y_j=\delta_{i,j}$ ($\partial x_i/\partial x_j$, in contrast, is not necessarily equal to $\delta_{i,j}$).   The relative two-fermion Green's function $G^{(\text{2F})}(E-E_\ell,z,z')$, e.g., depends on three variables and the expression $\partial_2 G^{(\text{2F})}(E-E_\ell,0,z')$ indicates that the derivative of the Green's function with respect to $z$ is taken for fixed $E-E_\ell$ and $z'$ and that $z$ is subsequently replaced by $0$. 

    Inserting the explicit form of $\hat{V}_{\text{int}}$ for three identical fermions into Eq.~(\ref{eq_ls_wavefunction}), we find  
        \begin{align}
        \label{eq_appendix_aux1}
             \psi(z_{12},z_{12,3}) &=g_{-}\intinf\intinf G(E,\mathbf{z},\mathbf{z}') \frac{\partial \delta(z_{12}')}{\partial z_{12}'}\frac{\partial \psi(z_{12}',z_{12,3}')}{\partial z_{12}'}~dz_{12}'dz_{12,3}' \nonumber \\
             &+g_{-} \intinf\intinf G(E,\mathbf{z},\mathbf{z}') \frac{\partial \delta(z_{13}')}{\partial z_{13}'}\frac{\partial \psi(z_{12}',z_{12,3}')}{\partial z_{13}'}~dz_{12}'dz_{12,3}' \nonumber \\
             &+g_{-}\intinf\intinf G(E,\mathbf{z},\mathbf{z}') \frac{\partial \delta(z_{23}')}{\partial z_{23}'}\frac{\partial \psi(z_{12}',z_{12,3}')}{\partial z_{23}'}~dz_{12}'dz_{12,3}',
        \end{align}
        where the relative three-fermion Green's function $G(E,\mathbf{z},\mathbf{z}')$ depends on five variables, namely $E$, $z_{12}$, $z_{12,3}$, $z_{12}'$, and $z_{12,3}'$.
        In writing Eq.~(\ref{eq_appendix_aux1}), we divided the right-hand side into three separate integrals since our aim in what follows is to transform the second and third integrals  to coordinates that  ``best match'' the respective delta-function in the integrand. Specifically, we 
    use the relations
        \begin{align}
                z_{12}=\frac{1}{2}z_{13}+\frac{\sqrt{3}}{2}z_{13,2}=-\frac{1}{2}z_{23}-\frac{\sqrt{3}}{2}z_{23,1} 
                \mbox{ and }
                z_{12,3}=\frac{\sqrt{3}}{2}z_{13}-\frac{1}{2}z_{13,2}=\frac{\sqrt{3}}{2}z_{23}-\frac{1}{2}z_{23,1}
        \end{align}
    to transform the second and third integrals on the right-hand side of Eq.~(\ref{eq_appendix_aux1}) to the coordinates \((z_{13}',z_{13,2}')\) and \((z_{23}',z_{23,1}')\), respectively:
        \begin{align}\label{somemoreints}
             \psi(z_{12},z_{12,3}) &=g_{-}\intinf\intinf G(E,\mathbf{z},\mathbf{z}') \frac{\partial \delta(z_{12}')}{\partial z_{12}'}\frac{\partial \psi(z_{12}',z_{12,3}')}{\partial z_{12}'}~dz_{12}'dz_{12,3}' \nonumber \\
             &+g_{-} \intinf\intinf G\lp E,\mathbf{z}, \frac{1}{2}z_{13}'+\frac{\sqrt{3}}{2}z_{13,2}', \frac{\sqrt{3}}{2}z_{13}'-\frac{1}{2}z_{13,2}'\rp \frac{\partial \delta(z_{13}')}{\partial z_{13'}}\frac{\partial \psi ( z_{12}',z_{12,3}')}{\partial z_{13}'}~dz_{13}'dz_{13,2}' \nonumber \\&
             +g_{-}\intinf\intinf G\lp E,\mathbf{z},-\frac{1}{2}z_{23}'-\frac{\sqrt{3}}{2}z_{23,1}', \frac{\sqrt{3}}{2}z_{23}'-\frac{1}{2}z_{23,1}'\rp \frac{\partial \delta(z_{23}')}{\partial z_{23}'}\frac{\partial \psi ( z_{12}',z_{12,3}')}{\partial z_{23}'}~dz_{23}'dz_{23,1}'.
        \end{align}
    Using that \(\psi\) must be   antisymmetric  under the exchange of two identical fermions, we obtain 
        \begin{align}
        \label{eq_antisymm}
            \psi(z_{13},z_{13,2})=-\psi(z_{12},z_{12,3}) \mbox{ and } \psi(z_{23},z_{23,1})=\psi(z_{12},z_{12,3}).
        \end{align}
    Substituting Eq.~(\ref{eq_antisymm})  into Eq. \eqref{somemoreints}, we find
        \begin{align}
        \label{eq_appendix_aux2}
             \psi(z_{12},z_{12,3}) &=g_{-}\intinf\intinf G(E,\mathbf{z},\mathbf{z}') \frac{\partial \delta(z_{12}')}{\partial z_{12}'}\frac{\partial \psi(z_{12}',z_{12,3}')}{\partial z_{12}'}~dz_{12}'dz_{12,3}' \nonumber \\
             &-g_{-} \intinf\intinf G\lp E,\mathbf{z}, \frac{1}{2}z_{13}'+\frac{\sqrt{3}}{2}z_{13,2}', \frac{\sqrt{3}}{2}z_{13}'-\frac{1}{2}z_{13,2}'\rp \frac{\partial \delta(z_{13}')}{\partial z_{13}'}\frac{\partial \psi ( z_{13}',z_{13,2}')}{\partial z_{13}'}~dz_{13}'dz_{13,2}' \nonumber \\
             &+g_{-}\intinf\intinf G\lp E,\mathbf{z},-\frac{1}{2}z_{23}'-\frac{\sqrt{3}}{2}z_{23,1}', \frac{\sqrt{3}}{2}z_{23}'-\frac{1}{2}z_{23,1}'\rp \frac{\partial \delta(z_{23}')}{\partial z_{23}'}\frac{\partial \psi ( z_{23}',z_{23,1}')}{\partial z_{23}'}~dz_{23}'dz_{23,1}'.
        \end{align}
        To simplify the notation, we denote the variables $z_{12}$ and $z_{12,3}$ by $x$ and $y$. Moreover, in all three integrals on the right-hand side of Eq.~(\ref{eq_appendix_aux2}), we denote the dummy variables that are being integrated over by $x'$ and $y'$. This yields 
        \begin{align}\label{xandy}
             \psi(x,y) &=g_{-}\intinf\intinf G(E,x,y,x',y') \frac{\partial \delta(x')}{\partial x'}\frac{\partial \psi(x',y')}{\partial x'}~dx' dy' \nonumber \\
             &-g_{-} \intinf\intinf G\lp E,x,y, \frac{1}{2}x'+\frac{\sqrt{3}}{2}y', \frac{\sqrt{3}}{2}x'-\frac{1}{2}y'\rp \frac{\partial \delta(x')}{\partial x'}\frac{\partial \psi\lp x',y'\rp}{\partial x'}~dx'dy' \nonumber \\
             &+g_{-}\intinf\intinf G\lp E,x,y,-\frac{1}{2}x'-\frac{\sqrt{3}}{2}y', \frac{\sqrt{3}}{2}x'-\frac{1}{2}y'\rp \frac{\partial \delta(x')}{\partial x'}\frac{\partial \psi\lp x',y'\rp}{\partial x'}~dx'dy'.
        \end{align}
    Applying the identity
        \begin{align}
            \int_{-\infty}^\infty f(x)\frac{\partial \delta(x)}{\partial x}dx=-\frac{\partial f(x)}{\partial x}\bigg|_{x=0},
        \end{align}
        which follows from integration by parts, to 
    Eq.~\eqref{xandy}, the two-dimensional integrals reduce to one-dimensional integrals:
        \begin{align}
            \psi(x,y) &=-g_{-}\intinf \lb\parder{x'}\lp G(E,x,y,x',y')\frac{\partial \psi(x',y')}{\partial x'}\rp\rb_{x'=0}~ dy' \nonumber \\
             &+g_{-} \intinf\lb\parder{x'}\lp G\lp E,x,y, \frac{1}{2}x'+\frac{\sqrt{3}}{2}y', \frac{\sqrt{3}}{2}x'-\frac{1}{2}y'\rp\frac{\partial \psi\lp x',y'\rp}{\partial x'}\rp\rb_{x'=0}~dy' \nonumber \\
             &-g_{-}\intinf\lb\parder{x'}\lp G\lp E,x,y,-\frac{1}{2}x'-\frac{\sqrt{3}}{2}y', \frac{\sqrt{3}}{2}x'-\frac{1}{2}y'\rp \frac{\partial \psi\lp x',y'\rp}{\partial x'}\rp\rb_{x'=0}~dy'.
        \end{align}
    Expanding the derivatives using the product and chain rules and employing the shorthand notation defined in Eq.~\eqref{shorthand}, we find
        \begin{align}\label{nowdiff}
            \psi(x,y) &=-g_{-}\intinf \lb \partial_4 G(E,x,y,0,y') \rb \partial_1 \psi(0,y')~ dy' \nonumber \\
            &+g_{-}\intinf \lb \frac{1}{2}\partial_4 G\lp E,x,y,\frac{\sqrt{3}}{2}y'
            ,-\frac{1}{2}y'\rp + \frac{\sqrt{3}}{2}\partial_5 G\lp E,x,y,\frac{\sqrt{3}}{2}y',-\frac{1}{2}y'\rp\rb\partial_1\psi(0,y')~dy' \nonumber \\
            &-g_{-}\intinf \lb -\frac{1}{2}\partial_4 G\lp E,x,y,-\frac{\sqrt{3}}{2}y'
            ,-\frac{1}{2}y'\rp + \frac{\sqrt{3}}{2}\partial_5 G\lp E,x,y,-\frac{\sqrt{3}}{2}y',-\frac{1}{2}y'\rp\rb\partial_1\psi(0,y')~dy'.
        \end{align}
        In Eq.~(\ref{nowdiff}), we use
 that \( \frac{\partial^2 \psi(x',y')}{\partial x'^2}\big|_{x'=0}\) is equal to zero due to the anti-symmetry requirement under the exchange of two identical fermions [i.e., $\psi$ is an odd function in $x'$, $\partial \psi(x',y')/\partial x'$ is an even function in $x'$, and $\partial^2 \psi(x',y')/\partial x'^2$ is an odd function in $x'$]. 
 
    To proceed, we differentiate both sides of Eq.~\eqref{nowdiff} with respect to \(x\) and subsequently evaluate the resulting expressions in the limit \(x=0^{+}\). Since the functions that are being evaluated at \(x=0^{+}\) are even in $x$, we  replace the limit \(x=0^{+}\)  by \(x=0\). This yields
        \begin{align}\label{toexpand}
            \partial_1\psi(0,y) &=-g_{-}\intinf [\partial_{2,4} G(E,0,y,0,y')]\partial_1 \psi(0,y')~ dy' \nonumber \\
            &+g_{-}\intinf \lb \frac{1}{2}\partial_{2,4} G\lp E,0,y,\frac{\sqrt{3}}{2}y'
            ,-\frac{1}{2}y'\rp + \frac{\sqrt{3}}{2}\partial_{2,5} G\lp E,0,y,\frac{\sqrt{3}}{2}y',-\frac{1}{2}y'\rp\rb\partial_1\psi(0,y')~dy' \nonumber \\
            &-g_{-}\intinf \lb -\frac{1}{2}\partial_{2,4} G\lp E,0,y,-\frac{\sqrt{3}}{2}y'
            ,-\frac{1}{2}y'\rp + \frac{\sqrt{3}}{2}\partial_{2,5} G\lp E,0,y,-\frac{\sqrt{3}}{2}y',-\frac{1}{2}y'\rp\rb\partial_1\psi(0,y')~dy'.
        \end{align}
    The right-hand side of Eq.~(\ref{toexpand}) contains two types of Green's function derivatives, namely,  $\partial_{2,4}G$ and $\partial_{2,5}G$. 
 To proceed, we rewrite each of these Green's function derivatives in terms of derivatives of the relative two-fermion Green's function  $G^{(\text{2F})}$. For $\partial_{2,4}G( E,0,y,0,y')$, we write
\begin{align}\label{g240}
            \partial_{2,4}G( E,0,y,0,y')=\sum_{i,j=0,1,2,\dots} \frac{\left[\frac{\partial \phi_i^{*}(x')}{\partial x'}\right]_{x'=0} \phi_j^*\lp y' \rp \left[\frac{\partial \phi_i(x)}{\partial x} \right]_{x=0}\phi_j\lp y\rp}{E_i+E_j-E},
            \end{align}
which can, after rearranging the sums and using the definition of the relative two-fermion Green's function [Eq.~(\ref{greenfuncnbody}) with $\mathbf{z}=x$],
be rewritten as 
            \begin{align}
            \label{g240final}
               \partial_{2,4}G( E,0,y,0,y')= 
            \sum_{j=0}^\infty \phi_j^*\lp y'\rp \phi_j\lp y\rp \partial_{2,3}G^{(\text{2F})}(E-E_j,0,0).
        \end{align}
        Similarly, one finds
  \begin{align}\label{g24}
            \partial_{2,4}G\lp E,0,y,\pm\frac{\sqrt{3}}{2}y',-\frac{1}{2}y'\rp= \sum_{j=0}^\infty \phi_j^*\lp -\frac{1}{2}y'\rp \phi_j\lp y\rp \partial_{2,3}G^{(\text{2F})}\lp E-E_j,0,\pm\frac{\sqrt{3}}{2}y' \rp
        \end{align}
    and 
        \begin{align}\label{g25}
            \partial_{2,5}G\lp E,0,y,\pm\frac{\sqrt{3}}{2}y',-\frac{1}{2}y'\rp=            
            \sum_{j=0}^\infty 
            \left[ \frac{\partial \phi_j^{*}(x')}{\partial x'}\right]_{x'=-y'/2} \phi_j\lp y\rp \partial_{2}G^{(\text{2F})}\lp E-E_j,0,\pm\frac{\sqrt{3}}{2}y' \rp.
        \end{align}
Using that $\partial_2G^{(\text{2F})}(E-E_j,0,y')$ is an odd function in $y'$, one readily obtains
        \begin{align}\label{g25_2}
            \partial_{2,5}G\lp E,0,y,\frac{\sqrt{3}}{2}y',-\frac{1}{2}y'\rp
            -
            \partial_{2,5}G\lp E,0,y,-\frac{\sqrt{3}}{2}y',-\frac{1}{2}y'\rp=\nonumber \\ 2           
            \sum_{j=0}^\infty 
            \left[ \frac{\partial \phi_j^{*}(x')}{\partial x'}\right]_{x'=-y'/2} \phi_j\lp y\rp \partial_{2}G^{(\text{2F})}\lp E-E_j,0,\frac{\sqrt{3}}{2}y' \rp.
        \end{align}
Since our ultimate goal is to rewrite the Green's functions (or their derivatives) in terms of an infinite series that contains harmonic oscillator functions (this is what is needed to make use of the recursion relationship for $W_{\bullet}$), the right-hand side of Eq.~(\ref{g24}) cannot be used directly in Eq.~(\ref{toexpand}); specifically,  $\partial_{2,3}G^{(\text{2F})}(E-E_j,0,\pm \sqrt{3}y'/2)$ is discontinuous at $y'=0$, preventing us from directly seeking a series expansion. 
To ``resolve'' this discontinuity, we break the integral over $\partial_{2,4}G$ into two pieces, 
      \begin{align}\label{g24int22}
            \intinf \partial_{2,4}  G\lp E,0,y,\pm\frac{\sqrt{3}}{2}y',-\frac{1}{2}y'\rp  f(y') dy'    
            =\int_{0^{+}}^\infty \lb \sum_{j=0}^\infty \phi_j^*\lp -\frac{1}{2}y'\rp \phi_j\lp y\rp \partial_{2,3}G^{(\text{2F})}\lp E-E_j,0,\pm\frac{\sqrt{3}}{2}y' \rp\rb f(y') dy' \nonumber \\
            + \int_{-\infty}^{0^{-}} \lb \sum_{j=0}^\infty \phi_j^*\lp -\frac{1}{2}y'\rp \phi_j\lp y\rp \partial_{2,3}G^{(\text{2F})}\lp E-E_j,0,\pm\frac{\sqrt{3}}{2}y' \rp\rb f(y') dy' ,
            \end{align}
            where we assume that $\parder{y'}\lb f(y')\sum_{j=0}^\infty \phi_j^*\lp -\frac{1}{2}y'\rp \phi_j\lp y\rp\rb$ is  continuous in $y'$ for all $y'$ except for possibly $y'=0$. 
            Integrating both integrals on the right-hand side of Eq.~(\ref{g24int22}) by parts, we find  
            \begin{align}
            \intinf \partial_{2,4} & G\lp E,0,y,\pm\frac{\sqrt{3}}{2}y',-\frac{1}{2}y'\rp  f(y')dy'=
            \nonumber \\
            & \lb\pm\frac{2}{\sqrt{3}}\partial_2 G^{(\text{2F})}\lp E-E_j,0,\pm\frac{\sqrt{3}}{2}y' \rp\lp\sum_{j=0}^\infty \phi_j^*\lp -\frac{1}{2}y'\rp \phi_j\lp y\rp\rp f(y')\rb\Bigg\vert_{0^{+}}^\infty \nonumber \\
            &\mp \frac{2}{\sqrt{3}}\int_{0^{+}}^\infty \partial_2 G^{(\text{2F})}\lp E-E_j,0,\pm\frac{\sqrt{3}}{2}y' \rp \lb\parder{y'}\lp f(y')\sum_{j=0}^\infty \phi_j^*\lp -\frac{1}{2}y'\rp \phi_j\lp y\rp\rp\rb~dy' \nonumber \\
            &+\lb\pm\frac{2}{\sqrt{3}}\partial_2 G^{(\text{2F})}\lp E-E_j,0,\pm\frac{\sqrt{3}}{2}y' \rp\lp\sum_{j=0}^\infty \phi_j^*\lp -\frac{1}{2}y'\rp \phi_j\lp y\rp\rp f(y')\rb\Bigg\vert_{-\infty}^{0^{-}} \nonumber \\
            &\mp \frac{2}{\sqrt{3}}\int_{-\infty}^{0^{-}} \partial_2 G^{(\text{2F})}\lp E-E_j,0,\pm\frac{\sqrt{3}}{2}y' \rp \lb\parder{y'}\lp f(y')\sum_{j=0}^\infty \phi_j^*\lp -\frac{1}{2}y'\rp \phi_j\lp y\rp\rp\rb dy'. \end{align}
            Recombining the two integrals and using \(\partial_2 G^{(\text{2F})}(E-E_j,0,0^{\pm})=\pm 1\), 
             we find
            \begin{align}
                \intinf \partial_{2,4}  G\lp E,0,y,\pm\frac{\sqrt{3}}{2}y',-\frac{1}{2}y'\rp  f(y')dy
            =-\frac{4}{\sqrt{3}}\lb \sum_{j=0}^\infty \phi_j^*\lp 0\rp\phi_j(y)\rb\lb f(0^{+})+f(0^{-})\rb \nonumber \\
            \mp \frac{2}{\sqrt{3}}\intinf \partial_2 G^{(\text{2F})}\lp E-E_j,0,\pm\frac{\sqrt{3}}{2}y' \rp \lb\parder{y'}\lp f(y')\sum_{j=0}^\infty \phi_j^*\lp -\frac{1}{2}y'\rp \phi_j\lp y\rp\rp\rb dy'.
        \end{align}
Using again that $\partial_2G^{(\text{2F})}(E-E_j,0,y')$ is an odd function in $y'$, 
one readily finds
\begin{align}\label{g24int2}
            \intinf \partial_{2,4} G\lp E,0,y,\frac{\sqrt{3}}{2}y',-\frac{1}{2}y'\rp f(y')~dy'
            +
           \intinf \partial_{2,4} G\lp E,0,y,-\frac{\sqrt{3}}{2}y',-\frac{1}{2}y'\rp f(y') dy'
           = \nonumber \\
   -\frac{8}{\sqrt{3}}\lb \sum_{j=0}^\infty \phi_j^*\lp 0\rp\phi_j(y)\rb\lb f(0^{+})+f(0^{-})\rb 
   - \nonumber \\
   \frac{4}{\sqrt{3}}\intinf \partial_2 G^{(\text{2F})}\lp E-E_j,0,\pm\frac{\sqrt{3}}{2}y' \rp \lb\parder{y'}\lp f(y')\sum_{j=0}^\infty \phi_j^*\lp -\frac{1}{2}y'\rp \phi_j\lp y\rp\rp\rb dy'.
            \end{align}

Substituting Eqs.~(\ref{g240final}), (\ref{g25_2}), and (\ref{g24int2}) into Eq.~(\ref{toexpand}), we find 
\begin{align}
\label{toexpand2}
\partial_1 \psi(0,y) 
             &=
            -g_{-} \intinf\lb \sum_{j=0}^\infty \phi_j^*\lp y'\rp \phi_j\lp y\rp \partial_{2,3}G^{(\text{2F})}\lp E-E_j,0,0\rp\rb  \partial_1 \psi(0,y')  dy' \nonumber \\
            &-\frac{2 g_{-}}{\sqrt{3}}\intinf\partial_2 G^{(\text{2F})}\lp E-E_j,0,\frac{\sqrt{3}}{2}y'\rp \parder{y'}\lb\sum_{j=0}^\infty \phi_j^*\lp -\frac{1}{2}y'\rp \phi_j(y) \partial_1 \psi(0,y') \rb dy' \nonumber \\
            &+\frac{2\sqrt{3}g_{-}}{2}\intinf \Bigg\{ \sum_{j=0}^\infty \left[ \frac{\partial \phi_j^{*}(x')}{\partial x'}\right]_{x'=-y'/2} \phi_j\lp y\rp \partial_{2}G^{(\text{2F})}\lp E-E_j,0,\frac{\sqrt{3}}{2}y' \rp \Bigg\} \partial_1 \psi(0,y') dy' \nonumber \\
            &-\frac{2g_{-}}{\sqrt{3}}\lb\sum_{j=0}^\infty \phi_j^*(0) \phi_j(y)\rb\lb 
            \partial_1 \psi(0,0^+)+\partial_1 \psi(0,0^-)
            \rb.
        \end{align}
    Our next step consists of expanding \(\partial_1\psi(0,y)\) on the left-hand side and in all integrands on the right-hand side of 
    Eq.~(\ref{toexpand2})  in terms of the complete set of harmonic oscillator states $\phi_n(y)$, 
        \begin{align}\label{expansion}
            \partial_1\psi(0,y)=\sum_{n=0}^\infty a_n\phi_n(y),
        \end{align}
     multiplying both sides of Eq.~(\ref{toexpand2}) by $\phi_\ell^*(y)$, and subsequently integrating the resulting equation over $y$. Using the orthogonality of the harmonic oscillator states and that  \(\phi_k(0^{+})+\phi_k(0^{-})\) is equal to \(2\phi_k(0)\) for all \(k=0,1,2,\dots\), we find 
        \begin{align}
            \frac{1}{g_{-}}a_{\ell} &=\sum_{k=0}^\infty a_{k}\bigg\{-\partial_{2,3}G^{(\text{2F})}(E-E_{\ell},0,0)\delta_{k,\ell} \nonumber \\
            &-\frac{2}{\sqrt{3}}\intinf\partial_2 G^{(\text{2F})}\lp E-E_\ell,0,\frac{\sqrt{3}}{2}y'\rp \parder{y'}\lb \phi_\ell^*\lp -\frac{1}{2}y'\rp \phi_k(y')\rb dy' \nonumber \\
            &+\sqrt{3}\intinf \left[ \frac{\partial \phi_j^{*}(x')}{\partial x'}\right]_{x'=-y'/2}\partial_{2}G^{(\text{2F})}\lp E-E_\ell,0,\frac{\sqrt{3}}{2}y' \rp \phi_k(y') dy' - \frac{4}{\sqrt{3}}\lb \phi_{\ell}^*(0)\phi_k(0)\rb \bigg\}.
        \end{align}
    Applying the product rule and 
    combining like terms, we finally arrive at 
        \begin{align}
            \frac{1}{g_{-}}a_{\ell} &=\sum_{k=0}^\infty a_{k}\bigg\{-\partial_{2,3}G^{(\text{2F})}(E-E_{\ell},0,0)\delta_{k,\ell} \nonumber \\
            &-\frac{2}{\sqrt{3}}\intinf \phi_{\ell}^*\lp -\frac{1}{2}y'\rp \frac{\partial\phi_k(y')}{\partial y'}\partial_2 G^{(\text{2F})}\lp E-E_\ell,0,\frac{\sqrt{3}}{2}y'\rp dy' \nonumber \\
            &+\frac{4}{\sqrt{3}}\intinf \left[ \frac{\partial \phi_\ell^{*}(x')}{\partial x'}\right]_{x'=-y'/2}\phi_k(y')\partial_{2}G^{(\text{2F})}\lp E-E_\ell,0,\frac{\sqrt{3}}{2}y' \rp dy' 
            - \frac{4}{\sqrt{3}}\lb \phi_{\ell}^*(0)\phi_k(0)\rb \bigg\},
        \end{align}
        which agrees with Eqs.~(\ref{goaleqfff})-(\ref{eq_integral_capk}) of the main text.


%
        
\end{document}